\documentclass[12pt,a4paper]{article}
\usepackage{amsfonts}
\usepackage{mathrsfs}  
\usepackage{epsfig}
\usepackage{axodraw}
\usepackage{graphicx}

\def\simlt{\stackrel{<}{{}_\sim}}
\def\simgt{\stackrel{>}{{}_\sim}}

\begin{document}

\title{Perturbative computation of thermal characteristics of the
  Stoner phase transition}
\author{\em Oskar Grocholski$~\!^1$ and Piotr H.
  Chankowski$~\!^2$\footnote{Emails:
    oskar.grocholski@desy.de, chank@fuw.edu.pl}\\
  $^1\phantom{a}$Deutsches Electronen-Synchrotron DESY\\
  Notkestr. 85, 22607 Hamburg, Germany\\
$^2$Faculty of Physics, University of Warsaw,\\
Pasteura 5, 02-093 Warsaw, Poland
}
\maketitle
\abstract{We  apply the thermal
  (imaginary time) perturbative expansion 
  to the relevant effective field theory to compute characteristics
  of the phase transition to the ordered state which
  can occur at low temperatures in the gas of (nonrelativistic) spin $1/2$
  fermions interacting through a short range spin independent repulsive
  binary interaction potential. We show how to obtain a systematic
  expansion of the system's free energy depending on the densities $n_+$
  and $n_-$ of spin up and spin down fermions. In this paper we truncate
  this expansion at the second order and determine, by numerically
  minimizing the free energy, the equilibrium proportions of $n_+$ and $n_-$
  (that is, the system's polarization) as functions of the temperature, the
  system's overall density $n=n_++n_-$ and the strength of the interaction. 
\vskip0.1cm

\noindent{\em Keywords}: Diluted gas of interacting fermions, phase
transitions, thermodynamic potentials, effective field
theory, scattering lengths}

\newpage


\section{Introduction}
\label{sec:introd}

There is a qualitative argument that in the gas of spin 1/2 fermions
interacting through a short range repulsive spin-independent binary potential 
a phase transition to the ordered state should occur, if the interaction
strength and/or the system's density is sufficiently large. Indeed, at zero
temperature, when the entropy factor does not intervene, the configuration
of the system in
which there are more fermions in one spin state than in the other one may be
energetically favoured because, due to the Pauli exclusion principle the
$s$-wave interaction of fermions in the same spin state is impossible and
the resulting decrease of the interaction energy may be greater than the
associated increase of the kinetic energy (increase of the Fermi energy of
the more populated spin state). 

Theoretical investigation of this phenomenon, called the Stoner transition,
taking into account its temperature dependence, requires the full machinery of
statistical mechanics. The standard textbook treatment of the problem
\cite{Kesio,Pathria}, equivalent to the so-called mean field approach or the
Hartree-Fock approximation, employs the pseudo-potential method which allows
to determine in the first order approximation the Hamiltonian spectrum and
to compute the Canonical Ensemble partition function of the
system. In this approximation the phase transition is continuous (with the
divergent magnetic susceptibility characterized by the critical exponent
$\gamma=1$ and a finite discontinuity of the heat capacity) and at low
temperatures (where the Sommerfeld expansion can be used to obtain analytical
expression for the relevant chemical potentials) it occurs when
\cite{Stoner,Kesio,Pathria}
\begin{eqnarray}
  k_{\rm F}a_0\geq{\pi\over2}\left[1+{\pi^2\over12}
    \left({k_{\rm B}T\over\varepsilon_{\rm F}}\right)^2+\dots\right],\nonumber
\end{eqnarray}
where the (overall) Fermi wave vector and energy
\begin{eqnarray}
  k_{\rm F}=\left(3\pi^2~\!{N\over V}\right)^{1/3},\phantom{aaaa}
  \varepsilon_{\rm F}={\hbar^2k_{\rm F}^2\over2m_f}~\!,\label{eqn:spin1/2defsofkFandEpsF}
\end{eqnarray}
characterize the density of the system and $a_0>0$ is the $s$-wave scattering
length characterizing the strength of the (repulsive) interaction. The
continuous character of the Stoner transition obtained in this approximation
is, however, accidental - it is due to a numerical coincidence specific for a
(three-dimensional) system of spin $s=1/2$ fermions only (in the same
approximation the transition is of first order if $s>1/2$ and/or $D\neq3$).
In fact, computing the system's free energy beyond the mean field approximation,
using the ordinary second order perturbative expansion, it was found
\cite{DUMacDO} that at low temperatures it is of the first order,
just as had been suggested in \cite{BeKiVo} on the basis of the
generic presence of non-analytic terms (resulting from the coupling of the
order parameter to the gap-less modes) in the free energy which 
cause the transition to have the first order character. 

The character of the considered transition (its dependence on the
parameter $k_{\rm F}a_0$) can be most easily investigated at
zero temperature because then the problem reduces to the computation of the
ground-state energy density $E_\Omega/V$ of the system of fermions
interacting through a binary spin-independent repulsive potential as a
function of the system's density $n=N/V$ and its polarization
\begin{eqnarray}
  P=(N_+-N_-)/N~\!.\label{eqn:defP}
\end{eqnarray}
Such a computation is most easily performed using the modern effective
field theory approach, the application of which to this problem  has been
pioneered in \cite{HamFur00}. In this approach the underlying spatially
nonlocal field-theory interaction (see e.g. \cite{FetWal}
for the exposition of the relevant formalism of the second quantization),
resulting from the ordinary potential two-body interaction, is replaced by
an infinite set of local (contact) effective interactions
\begin{eqnarray}
\hat V_{\rm int}=C_0\!\int\!d^3\mathbf{x}~\!\psi_+^\dagger\psi_+\psi_-^\dagger\psi_-
+\hat V_{\rm int}^{(C_2)}+V_{\rm int}^{(C_2^\prime)}+\dots  \label{eqn:Veff}
\end{eqnarray}
$\psi_\pm(\mathbf{x})$ are here the usual field operators of spin up and
down fermions; the terms $V_{\rm int}^{(C_2)}$, $V_{\rm int}^{(C_2^\prime)}$
(which will be not needed in this work) represent local operators
of lower length dimension with four fermionic fields and two spatial
derivatives and the ellipsis stands for other local operators (with
more derivatives and/or field operators) of yet
lower length dimension (see \cite{HamFur00}). The amount of work
needed to obtain the systematic expansion of the ground state energy in
powers (which can be modified by logarithms) of $k_{\rm F}R$, where $R$
is the characteristic length scale of the underlying two-body spin
independent interaction potential, is in this way greatly reduced. This
is because in this approach the coupling constants, like $C_0$ in
(\ref{eqn:Veff}), of the effective local interactions are directly
determined in terms of the scattering lengths $a_\ell$ and effective radii
$r_\ell$, $\ell=0,1,\dots,$ (which are assumed to be of order $\sim R$)
parametrizing the low energy expansion in powers of the relative momentum
$\hbar|\mathbf{k}|$ of the elastic scattering amplitude of two fermions.
The simplifications brought in by the effective field theory method allowed
to easily reproduce \cite{CHWO1} and generalize to arbitrary repulsive
potentials and arbitrary spins $s$ \cite{PECABO} the old result of Kanno
\cite{KANNO} who computed the order $(k_{\rm F}R)^2$ correction to the energy
density using the specific hard sphere interaction of spin $s=1/2$ fermions.
The first order character
of the phase transition at $T=0$ is then clearly seen in the form of the
energy density obtained in this approximation plotted as a function of the
order parameter $P$: starting from some value of $k_{\rm F}a_0$ the energy
density develops the second minimum well away from the one at $P=0$ and
at $k_{\rm F}a_0=1.054$ (for $s=1/2$) this second minimum becomes deeper than
that at $P=0$.

However the analysis of the dependence on the order parameter of the system's
energy density which includes the complete order $(k_{\rm F}R)^3$ corrections
obtained recently in \cite{CHWO3,CHWO4} using the same effective field theory
approach shows that, independently of the value $s$ of the fermion spin, they
have the effect of erasing the first order character of the
Stoner transition, making it almost indistinguishable
from the continuous one. This is reflected in the fact that the height
of the hill separating the minimum at $P\neq0$ from the one at $P=0$
is greatly reduced (for higher spins also the position of the nontrivial
minimum of $E_\Omega/V$ as a function of the relevant order parameter
is strongly shifted towards its small values) compared to the situation
without these corrections. Moreover, there are claims \cite{He} based
on a resummation of an infinite subclass of Feynman diagrams
contributing to the ground-state energy density that the transition
(at $T=0$) is indeed continuous. Although it is not obvious that the
contributions taken into account in this resummation are really
the dominant ones \cite{CHWO4}, the results it leads to seem to agree well,
as far as the critical value of $k_{\rm F}a_0$ is concerned, with the
numerical quantum Monte Carlo simulations \cite{QMC10}.

In view of this situation it is desirable to investigate how the higher
order corrections influence the character of the Stoner phase transition
at nonzero temperatures. With this goal in mind in this paper we
formulate a systematic perturbative expansion of the thermodynamic
potentials of the system in question applying the standard imaginary time
formalism \cite{FetWal} within the effective field theory. We show that
the expansion of the free energy is in this approach particularly simple
being given by the same connected vacuum Feynman diagrams which give
nonzero contributions to the energy density expressed in terms of the
chemical potentials of the noninteracting system.
In the numerical analysis we restrict ourselves in this paper only to
the second order contributions reproducing the results obtained in
\cite{DUMacDO}, but with more labour the computations can be
extended to higher orders as well.

\section{Perturbative expansion of the thermodynamic potential $\Omega(T,V,\mu_+,\mu_-)$}

The natural equilibrium statistical physics formalism in which to treat
the problem of the gas of fermions the interactions of which preserve their
spins, and therefore the numbers $N_\sigma$ of particles with the spin projection
$\sigma$, is the Grand Canonical Ensemble with separate chemical potentials
$\mu_\sigma$ associated with the individual spin projections. One is therefore
interested in the statistical operator (as usually, $\beta\equiv1/k_{\rm B}T$)
\begin{eqnarray}
  \hat\rho={1\over\Xi_{\rm stat}}~\!e^{-\beta\hat K}~\!,\phantom{aaa}{\rm in~which}
  \phantom{aaa}\hat K=\hat H_0-\sum_\sigma\mu_\sigma\hat N_\sigma+\hat V_{\rm int}
  \equiv \hat K_0+\hat V_{\rm int}~\!,
\end{eqnarray}
and in computing the statistical sum (we specify the notation to the case of
spin $1/2$ fermions, so that $\sigma=+,-$)
\begin{eqnarray}
  \Xi_{\rm stat}(T,V,\mu_+,\mu_-)={\rm Tr}\!\left(e^{-\beta\hat K}\right),
\end{eqnarray}
from which all the necessary thermodynamic potentials can in principle
be obtained by performing the standard steps.
The free part $\hat K_0$ of the operator $\hat K=\hat K_0+\hat V_{\rm int}$,
where $\hat V_{\rm int}$ will be taken in the form (\ref{eqn:Veff}), is
\begin{eqnarray}
  \hat K_0=\sum_{\mathbf{p},\sigma}\left(\varepsilon_{\mathbf{p}}-\mu_\sigma\right)
  a_{\mathbf{p},\sigma}^\dagger
  a_{\mathbf{p},\sigma}=\sum_\sigma\int\!{d^3\mathbf{p}\over(2\pi)^3}
  \left(\varepsilon_{\mathbf{p}}-\mu_\sigma\right)
  a^\dagger_\sigma(\mathbf{p})~\!a_\sigma(\mathbf{p})~\!,
  \label{eqn:Kamiltonian0}
\end{eqnarray}
with $\varepsilon_{\mathbf{p}}\equiv\hbar^2\mathbf{p}^2/2m_f$,
in the normalizations in the finite volume $V$ and in an infinite
space, respectively.

To compute perturbatively the statistical sum $\Xi_{\rm stat}(T,V,\mu_+,\mu_-)$ one
introduces \cite{FetWal} the (imaginary time) interaction picture evolution operator
\begin{eqnarray}
  {\cal U}_I(\tau_2,\tau_1)=e^{\tau_2\hat K_0}~\!
  e^{-(\tau_2-\tau_1)\hat K}~\!e^{-\tau_1\hat K_0}~\!,
\end{eqnarray}
which satisfies the differential equation
\begin{eqnarray}
{d\over d\tau_2}~\!{\cal U}_I(\tau_2,\tau_1)=-V^I_{\rm int}(\tau_2)~\!
{\cal U}_I(\tau_2,\tau_1)~\!,\nonumber
\end{eqnarray}
($V^I_{\rm int}(\tau_2)=e^{\tau_2\hat K_0}V_{\rm int}e^{-\tau_2\hat K_0}$)
with the ``initial'' condition ${\cal U}_I(\tau,\tau)=\hat1$
and which formally can be written in the form
\begin{eqnarray}
  {\cal U}_I(\tau_2,\tau_1)={\rm T}_\tau
  \exp\!\left\{-\int_{\tau_1}^{\tau_2}\!d\tau~\!
  V^I_{\rm int}(\tau)\right\},\nonumber
\end{eqnarray}
in which ${\rm T}_\tau$ is the symbol of the ``chronological'' ordering.
Since $e^{-\beta\hat K}=e^{-\beta\hat K_0}~\!{\cal U}_I(\beta,0)$, the
statistical sum can be represented as
\begin{eqnarray}
  \Xi_{\rm stat}={\rm Tr}\!\left(e^{-\beta\hat K_0}~\!{\cal U}_I(\beta,0)\right)
  \equiv\Xi_{\rm stat}^{(0)}{\rm Tr}\!\left(\hat\rho^{(0)}~\!
           {\cal U}_I(\beta,0)\right),
\end{eqnarray}
where $\hat\rho^{(0)}$ and $\Xi_{\rm stat}^{(0)}$ are the statistical operator
and the statistical sum of the noninteracting system, respectively.
The perturbative expansion of $\Xi_{\rm stat}$ is then given by the series
\begin{eqnarray}
  \Xi_{\rm stat}=\Xi_{\rm stat}^{(0)}\sum_{n=0}^\infty{(-1)^n\over n!}\!
  \int_0^\beta\!d\tau_n\dots\int_0^\beta\!d\tau_1{\rm Tr}\left(\hat\rho^{(0)}~\!
{\rm T}_\tau[V^I_{\rm int}(\tau_n)\dots V^I_{\rm int}(\tau_1)]\right).
\end{eqnarray}
The corresponding expansion of the potential
$\Omega(T,V,\mu_+,\mu_-)=-{1\over\beta}\ln\Xi_{\rm stat}(T,V,\mu_+,\mu_-)$ is
\begin{eqnarray}
\Omega=\Omega^{(0)}-{1\over\beta}\ln\!\left\{\sum_{n=0}^\infty{(-1)^n\over n!}\!
  \int_0^\beta\!d\tau_n\dots\int_0^\beta\!d\tau_1{\rm Tr}\!\left(\hat\rho^{(0)}~\!
      {\rm T}_\tau[V^I_{\rm int}(\tau_n)\dots V^I_{\rm int}(\tau_1)]\right)\right\},
\end{eqnarray}
its first term $\Omega^{(0)}$ being the textbook expression \cite{Kesio}
($\varepsilon_{\mathbf{k}}=\hbar^2\mathbf{k}^2/2m_f$)
\begin{eqnarray}
  \Omega^{(0)}(T,V,\mu_\sigma)=-{1\over\beta}~\!
  \sum_\sigma V\!\int\!{d^3\mathbf{k}\over(2\pi)^3}\ln\!\left(
  1+e^{-\beta(\varepsilon_{\mathbf{k}}-\mu_\sigma)}\right).\label{eqn:Omega0}
\end{eqnarray}
Or, since the logarithm picks up connected contributions only,
\begin{eqnarray}
\Omega=\Omega^{(0)}-{1\over\beta}\sum_{n=0}^\infty{(-1)^n\over n!}\!
  \int_0^\beta\!d\tau_n\dots\int_0^\beta\!d\tau_1{\rm Tr}\!\left(\hat\rho^{(0)}~\!
      {\rm T}_\tau[V^I_{\rm int}(\tau_n)\dots V^I_{\rm int}(\tau_1)]\right)^{\rm con}.
      \label{eqn:OmegaExpansionInConnected}
\end{eqnarray}
In this form the expression for $\Omega$ is just the thermal analog of
the expansion of the formula\footnote{Here $T$ denotes time and not the
  temperature.}
\begin{eqnarray}
  E_\Omega=E_{\Omega_0}-\lim_{T\rightarrow\infty}{\hbar\over iT}~\!
  \langle\Omega_0|{\rm T}_t\exp\!\left(-{i\over\hbar}\!\int_{-T/2}^{T/2}\!dt~\!
  V^I_{\rm int}(t)\right)|\Omega_0\rangle^{\rm con}~\!,\label{eqn:EOmegaFormula}
\end{eqnarray}
used in \cite{HamFur00,CHWO1,CHWO3,CHWO4} for computing the ground state
energy $E_\Omega$ of the system. It is clear that the correspondence
between the two formalisms is $\beta\leftrightarrow iT/\hbar$ (it transforms
the $K$-picture operators into the Heisenberg picture ones and vice versa).
The formula (\ref{eqn:EOmegaFormula}) for the ground state energy is thus
obtained from the thermal expansion (\ref{eqn:OmegaExpansionInConnected})
by taking the limit $\beta\rightarrow\infty$ and simultaneously adjusting
the chemical potential $\mu_\sigma$ so that there are $N_\sigma$ particles
with the spin projection $\sigma$ (see below). 

Evaluation of the successive terms of the expansion 
(\ref{eqn:OmegaExpansionInConnected}) reduces, owing to
the thermal analog of the Wick formula (see \cite{FetWal}), to
drawing all possible connected Feynman diagrams with a given numbers
of different interaction vertices arising from $\hat V_{\rm int}$ joined
by the oriented lines and integrating over the positions $\mathbf{x}$ and
``times'' $\tau$ ascribed to these vertices the corresponding products of
free thermal propagators
\begin{eqnarray}
-{\cal G}^{(0)}_{\sigma_2,\sigma_1}(\tau_2-\tau_1;\mathbf{x}_2-\mathbf{x}_1)
={1\over\beta}\sum_n\int\!{d^3\mathbf{k}\over(2\pi)^3}~\!
e^{-i\omega^F_n(\tau_2-\tau_1)}~\!e^{i\mathbf{k}\cdot(\mathbf{x}_2-\mathbf{x}_1)}
\left(-\tilde{\cal G}^{(0)}_{\sigma_2,\sigma_1}(\omega^F_n,\mathbf{k})\right),
\nonumber
\end{eqnarray}
the Fourier transforms
$-\tilde{\cal G}^{(0)}_{\sigma_2,\sigma_1}$ of which have the form \cite{FetWal}
(the definition of $\omega^F_n$ as well as of $\omega^B_n$ are given in
(\ref{eqn:S1F}) and (\ref{eqn:S1B}))
\begin{eqnarray}
  -\tilde{\cal G}^{(0)}_{\sigma_2,\sigma_1}(\omega^F_n,\mathbf{k})
  ={-\delta_{\sigma_2,\sigma_1}\over i\omega^F_n
    -(\varepsilon_{\mathbf{k}}-\mu_{\sigma_1})}~\!,\nonumber
\end{eqnarray}
associated with the (oriented) lines connecting the vertices of the
diagram. The resulting Feynman rules in the ``momentum'' space are almost
identical with the ordinary ones except that the integrations over
frequencies are replaced by summations over the (fermionic) Matsubara
frequencies $\omega^F_n=(\pi/\beta)(2n+1)$, $n\in\mathbb{Z}$.
In this way one obtains the expansion of the potential $\Omega(T,V,\mu_+,\mu_-)$
the successive terms of which depend on the chemical potentials $\mu_+$ and
$\mu_-$ which must be adjusted in successive orders of the expansion to yield
through the relations
\begin{eqnarray}
  N_\pm=-\left(\partial\Omega/\partial\mu_\pm\right)_{T,V}~\!,\label{eqn:eNy}
\end{eqnarray}
the prescribed densities
$n_+=N_+/V$ and $n_-=N_-/V$ of particles with the spin projections up and down.
\vskip0.2cm

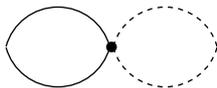
\begin{figure}[]
\begin{center}
\begin{picture}(80,40)(5,0)
\CArc(20,35)(20,195,345)
\CArc(20,25)(20,15,165)
\DashCArc(60,35)(20,195,345){2}
\DashCArc(60,25)(20,15,165){2}
\Vertex(40,30){2}
%
\end{picture}
\end{center}
\caption{The  first order correction $\Omega^{(1)}$ to the thermodynamic
  potential $\Omega(T,V,\mu_+,\mu_-)$. Solid and dashed lines represent
  fermions with opposite spin projections.}
\label{fig:VacuumDiagrSpinOneHalf}
\end{figure}

It will be instructive to recover first, using this formalism, the textbook results
\cite{Kesio,Pathria} of the mean field approximation. The first correction
$\Omega^{(1)}$ to the Grand potential is given by the single diagram shown in
Figure \ref{fig:VacuumDiagrSpinOneHalf}. The corresponding expression reads
\begin{eqnarray}
  \Omega^{(1)}={1\over\beta}~\!C_0\!\int_0^\beta\!d\tau\!\int\!d^3\mathbf{x}~\!
        {\rm Tr}\!\left(\hat\rho^{(0)}{\rm T}_\tau
        [\hat\psi^{\dagger I}_+\hat\psi^I_+\hat\psi^{\dagger I}_-\hat\psi^I_-]\right)
        =C_0~\!V~\!{\cal G}^{(0)}_{++}(0,\mathbf{0})
        ~\!{\cal G}^{(0)}_{--}(0,\mathbf{0})~\!.\label{eqn:Omega1}
\end{eqnarray}
Using the summation formula (\ref{eqn:S1F}) one obtains
\begin{eqnarray}
  {\cal G}^{(0)}_{\pm\pm}(0,\mathbf{0})
  =\int\!{d^3\mathbf{k}\over(2\pi)^3}
  \left[1+e^{\beta(\varepsilon_{\mathbf{k}}-\mu_\pm)}\right]^{-1}.\phantom{aaaaaaaa}
  \label{eqn:G(00)}
\end{eqnarray}

As will be shown in the next section, to the first order in the coupling
$C_0$ the free energy $F(T,V,N_+,N_-)$ is given by
\begin{eqnarray}
  F(T,V,N_+,N_-)=\Omega^{(0)}(T,V,\mu^{(0)}_+,\mu^{(0)}_-)
  +N_+\mu^{(0)}_++N_-\mu^{(0)}_-\nonumber\\
  +~\!\Omega^{(1)}(T,V,\mu^{(0)}_+,\mu^{(0)}_-)+\dots,
  \phantom{aaaaaaaaaaa}\!\label{eqn:FtoFirstOrderGeneral}
\end{eqnarray}
where $\mu^{(0)}_\pm$ are the zeroth order chemical potentials determined by
the conditions analogous to (\ref{eqn:eNy}) but with $\Omega$ replaced by
$\Omega^{(0)}$ given by (\ref{eqn:Omega0}). It is convenient to define the
function
\begin{eqnarray}
  f(\nu)\equiv{3\over2}\!\int_0^\infty\!d\xi~\!{\xi^{1/2}\over1+e^{\xi-\nu}}
  \equiv{3\sqrt\pi\over4}~\!f_{3/2}(\nu)~\!,\label{eqn:f32}
\end{eqnarray}
and to rewrite these conditions in the form
\begin{eqnarray}
  f(\nu_\pm)=\left({\varepsilon^{(0)}_{\rm F}(n_\pm)\over k_{\rm B}T}\right)^{3/2}~\!,
\end{eqnarray}
in which $\nu_\pm\equiv\mu_\pm/k_{\rm B}T$ and
$\varepsilon^{(0)}_{\rm F}(n)=(6\pi^2n)^{2/3}\hbar^2/2m_f$ is the Fermi
energy of the system of $N=nV$ spin $0$ noninteracting fermions enclosed
in the volume $V$. The function $f(\nu)$, which is a decent monotonically
growing function of $\nu$ mapping $\mathbb{R}$ onto $\mathbb{R}_+$,
has the inverse, so after writing $n_\pm$ as $(n/2)(1\pm P)$ the solutions
take the form\footnote{Inverting the appropriate expansions of the integral
  (\ref{eqn:f32}) given e.g. in \cite{Kesio} it is straightforward to
  find that asymptotically
  \begin{eqnarray}
    f^{-1}(x)=\left\{
    \matrix{\ln\!\left(\sqrt2-\sqrt{2-(4x/3)\sqrt{8/\pi}}\right)+\dots,
      \phantom{aa}x\ll1\cr
      x^{2/3}\left[1-(\pi^2/12)~\!x^{-4/3}-(\pi^4/80)~\!x^{-8/3}
        -(1511\pi^6/207360)~\!x^{-4}+\dots\right],
      \phantom{aa}x\gg1}\right.\nonumber
  \end{eqnarray}
}
\begin{eqnarray}
  {\mu^{(0)}_\pm\over k_{\rm B}T}=f^{-1}\!
  \left((1\pm P)\left({\varepsilon_{\rm F}(n)\over k_{\rm B}T}\right)^{3/2}\right),
  \label{eqn:nuDetermination}
\end{eqnarray}
in which $\varepsilon_{\rm F}(n)$ the system's overall Fermi energy
(\ref{eqn:spin1/2defsofkFandEpsF}).

Expressed in terms of the zeroth order chemical potentials $\mu_\pm^{(0)}$,
the first order correction (\ref{eqn:Omega1}) can be simply written as
$\Omega^{(1)}(T,V,\mu^{(0)}_+,\mu^{(0)}_-)=C_0V(N_+/V)(N_-/V)$, i.e. it is
independent (when expressed in terms of the particle densities) of the
temperature. Minimization with respect to $N_+$ of $F(T,V,N_+,N-N_+)$
truncated to the first order in the coupling $C_0$ (at fixed $N$) then
leads to the equilibrium condition
\begin{eqnarray}
\mu^{(0)}_+(N_+)-\mu^{(0)}_-(N-N_+)+{C_0\over V}~\!(N-2N_+)=0~\!,\nonumber
\end{eqnarray}
which, because $N-2N_+=-NP$ and (to this order) $C_0=(4\pi\hbar^2/m_f)a_0$, can
be rewritten in the familiar form \cite{Kesio}
\begin{eqnarray}
  \mu^{(0)}_+(N_+)-\mu^{(0)}_-(N_-)={8\over3\pi}~\!\varepsilon_{\rm F}~\!
  (k_{\rm F}a_0)~\!P~\!.\label{eqn:KesioCond}
\end{eqnarray}
This leads to the continuous phase transition.

The effect of the external magnetic field ${\cal H}$ can be also taken
into account by simply including the interaction with it in the free part
of the Hamiltonian, i.e. by replacing $\mu_\pm$ in (\ref{eqn:Kamiltonian0})
by $\tilde\mu_\pm=\mu_\pm\pm{\cal H}$
(the magnetic moment has been here included in ${\cal H}$ which has
therefore the dimension of energy). Since ultimately the free energy will
be cast in the form in which its dependence on $N_\pm$ and ${\cal H}$
enters only through $\tilde\mu^{(0)}_\pm$ which should be
determined from the conditions
\begin{eqnarray}
  {\tilde\mu^{(0)}_\pm\over k_{\rm B}T}={\mu^{(0)}_\pm\pm{\cal H}\over k_{\rm B}T}
  =f^{-1}\!\left((1\pm P)\left({\varepsilon_{\rm F}(n)\over k_{\rm B}T}\right)^{3/2}\right),
  \label{eqn:mu0withH}
\end{eqnarray}
this prescription remains valid to all orders of the expansion. In
particular, in the first order approximation the equilibrium condition,
written in the convenient dimensionless variables
\begin{eqnarray}
  t\equiv{T\over T_{\rm F}}\equiv{k_{\rm B}T\over\varepsilon_{\rm F}}~\!,\phantom{aaa}
  h\equiv{{\cal H}\over\varepsilon_{\rm F}}~\!,\phantom{aaa}
  \delta_\pm\equiv{\mu^{(0)}_\pm\over\varepsilon_{\rm F}}~\!,
  \label{eqn:dimlessvariables}
\end{eqnarray}  
takes the form
\begin{eqnarray}
  {8\over3\pi}~\!(k_{\rm F}a_0)~\!P+2h=t\left[
  f^{-1}\!\left({1+P\over t^{3/2}}\right)
  -f^{-1}\!\left({1-P\over t^{3/2}}\right)\right].
\end{eqnarray}
If the asymptotic expansion of $f^{-1}(x)$ for $x\gg1$ is used, this reproduces
the equilibrium condition derived in \cite{Kesio}.

For further applications it will be convenient to write down explicitly the
formula (\ref{eqn:FtoFirstOrderGeneral}) (including the external magnetic
field ${\cal H}$)
expressing it through the introduced dimensionless variables
(\ref{eqn:dimlessvariables}) and the polarization (\ref{eqn:defP}):
\begin{eqnarray}
  {6\pi^2\over k_{\rm F}^3}~\!{F\over\varepsilon_{\rm F}V}=
 -{3\sqrt\pi\over4}~\!t^{5/2}\left[f_{5/2}(\tilde\nu_+)
   +f_{5/2}(\tilde\nu_-)\right]
 \phantom{aaaaaaaaaaaaaaaaaaaaaaaaa}\nonumber\\
  +~\!(1+P)(\tilde\delta_+-h)
  +(1-P)(\tilde\delta_-+h)+(k_{\rm F}a_0)~\!{4\over3\pi}~\!(1-P^2)
    +\dots\label{eqn:FtoFirstOrderDimless}
\end{eqnarray}
Here\footnote{By the appropriate change of variables and the
  integration by parts $\Omega^{(0)}$ given by (\ref{eqn:Omega0})
  is written in terms of the standard integral (\ref{eqn:f52}) with $p=5/2$
  \cite{Kesio}.}
\begin{eqnarray}
  f_p(\nu)={1\over\Gamma(p)}\int_0^\infty\!{d\xi~\!\xi^{p-1}\over1+e^{\xi-\nu}}~\!,
  \label{eqn:f52}
\end{eqnarray}
and $\tilde\nu_\pm$ (and $\tilde\delta_\pm\equiv t\tilde\nu_\pm$) are given by
(\ref{eqn:mu0withH}).
In the limit $T\rightarrow0$ ($t\rightarrow0$) in which $\tilde\nu_\pm\gg1$,
$f_{5/2}(\nu)=(4/3\sqrt\pi)(2/5)\nu^{5/2}+\dots,$ while (c.f. the expansion
of the function $f^{-1}(x)$ given in the footnote)
$\tilde\nu_\pm=(1\pm P)^{2/3}+\dots$ and the right hand side of
(\ref{eqn:FtoFirstOrderDimless}) tends to
\begin{eqnarray}
  -{2\over5}\left[(1+P)^{5/3}+(1-P)^{5/3}\right]+
  (1+P)\left[(1+P)^{2/3}-h\right]\phantom{aaaaaaaaaaaaaaaaaa}~\!\nonumber\\
  + (1-P)\left[(1-P)^{2/3}+h\right]+(k_{\rm F}a_0)~\!{4\over3\pi}~\!(1-P^2)~\!,\nonumber
\end{eqnarray}
reproducing, of course, the well known formula for the ground-state energy
given (for ${\cal H}=0$) e.g. in \cite{CHWO4}.

\section{Expansion of the free energy}

From the thermodynamic point of view much more convenient to work with 
than the potential $\Omega$ is the free energy $F=\Omega+\mu_+N_++\mu_-N_-$
which canonically depends on $T$, $V$ and the particle numbers $N_\pm$.
It turns out that the expansion of this potential is also simpler. We will 
derive it here up to the third order following the method
outlined in \cite{FurHamPug}. To make the notation more transparent we
will denote the chemical potentials as
\begin{eqnarray}
  \mu_+\equiv x=x_0+x_1+x_2+\dots,\phantom{aaa}\mu_-\equiv y=y_0+y_1+y_2+\dots,
\end{eqnarray}
where the successive terms $x_n$, $y_n$ correspond to the successive terms
$\Omega^{(n)}$ of the expansion of the potential $\Omega$. Introducing the
notation $\Omega^{(n)}_x$, $\Omega^{(n)}_y$, $\Omega^{(n)}_{xx}$, etc. for the
first, second, etc. derivatives of $\Omega^{(n)}$ with respect to their
chemical potential arguments and expanding the right hand side of the relation
($N_x\equiv N_+$, $N_y\equiv N_-$)
\begin{eqnarray}
  F=\Omega^{(0)}(x_0+x_1+x_2+x_3+\dots,~\!y_0+y_1+y_2+y_3+\dots)
  \phantom{aaaaaaaaa}~\!\nonumber\\
  +~\!\Omega^{(1)}(x_0+x_1+x_2+\dots,~\!y_0+y_1+y_2+\dots)
  \phantom{aaaaaaaaaaaaaaaaa}~\nonumber\\
  +~\!\Omega^{(2)}(x_0+x_1+\dots,~\!y_0+y_1+\dots)
  +\Omega^{(3)}(x_0+\dots,y_0+\dots)+\dots\nonumber\\
  +~\!(x_0+x_1+x_2+x_3+\dots)N_x+(y_0+y_1+y_2+y_3+\dots)N_y~\!,\phantom{aaaa}~\! \nonumber
\end{eqnarray}
one obtains, using the zeroth order relations $\Omega^{(0)}_x=-N_x$ and
$\Omega^{(0)}_y=-N_y$ and the fact that
$\Omega^{(0)}(x_0,y_0)=\Omega^{\rm free}(x_0)+\Omega^{\rm free}(y_0)$
(cf. the formula (\ref{eqn:Omega0})), i.e. that $\Omega^{(0)}_{xy}=0$,
\begin{eqnarray}
  F=\left(\Omega^{(0)}+x_0N_x+y_0N_y\right)+\left(\Omega^{(1)}\right)
  \phantom{aaaaaaaaaaaaaaaaaaaaaaaaaaaaaaaaaaaa}\nonumber\\
  +\left(\Omega^{(2)}+x_1~\!\Omega^{(1)}_x+y_1~\!\Omega^{(1)}_y
  +{1\over2}~\!x_1^2~\!\Omega^{(0)}_{xx}
  +{1\over2}~\!y_1^2~\!\Omega^{(0)}_{yy}\right)
  \phantom{aaaaaaaaaaaaaaaaaaaaaa}\nonumber\\
  +\left(\Omega^{(3)}+x_1~\!\Omega^{(2)}_x+y_1~\!\Omega^{(2)}_y
  +{1\over2}~\!x_1^2~\!\Omega^{(1)}_{xx}
  +{1\over2}~\!y_1^2~\!\Omega^{(1)}_{yy}+x_1~\!y_1~\!\Omega^{(1)}_{xy}
  \right.\phantom{aaaaaaaaaaaaa}\label{eqn:F}\\
  \left.+~\!x_2~\!\Omega^{(1)}_x+y_2~\!\Omega^{(1)}_y
  +x_1~\!x_2~\!\Omega^{(0)}_{xx}+y_1~\!y_2~\!\Omega^{(0)}_{yy}
  +{1\over6}~\!x_1^3~\!\Omega^{(0)}_{xxx}+{1\over6}~\!y_1^3~\!\Omega^{(0)}_{yyy}
  \right)+\dots,\nonumber
\end{eqnarray}
all functions being now evaluated at $x_0$ and $y_0$ (at
$\tilde x_0=\mu_+^{(0)}+{\cal H}$ and $\tilde y_0=\mu_-^{(0)}-{\cal H}$
if there is an external magnetic field). The terms in the successive
brackets are the successive terms of the expansion of the free energy.
The first order correction $F^{(1)}$ used in the preceding section is
indeed given by $\Omega^{(1)}(x_0,y_0)$ (by $\Omega^{(1)}(\tilde x_0,\tilde y_0)$).
Furthermore, expanding around $x_0$ and $y_0$ (or $\tilde x_0$ and
$\tilde y_0$) the right hand side of the
relation (\ref{eqn:eNy}) which determines the chemical potential $x$
\begin{eqnarray}
  -N_x=\Omega^{(0)}_x+(x_1+x_2)~\!\Omega^{(0)}_{xx}+{1\over2}~\!x_1^2~\!\Omega^{(0)}_{xxx}
  +\Omega^{(1)}_x+x_1~\!\Omega^{(1)}_{xx}
  +y_1~\!\Omega^{(1)}_{xy}+\Omega^{(2)}_x+\dots,\nonumber
\end{eqnarray}
and the other similar relation for $y$, and taking into account that
$x_0$ and $y_0$ are such that $-N_x=\Omega^{(0)}_x$, $-N_y=\Omega^{(0)}_y$,
one obtains
\begin{eqnarray}
  x_1=-{\Omega^{(1)}_x\over\Omega^{(0)}_{xx}}~\!,
  \phantom{aaaaaaaaaaaaaaaaaaaaaaaaaaaaaa}~\!\nonumber\\
  x_2=-{\Omega^{(2)}_x\over\Omega^{(0)}_{xx}}
  +{\Omega^{(1)}_{xx}\Omega^{(1)}_x\over[\Omega^{(0)}_{xx}]^2}
  +{\Omega^{(1)}_{xy}\Omega^{(1)}_y\over\Omega^{(0)}_{xx}\Omega^{(0)}_{yy}}
  -{\Omega^{(0)}_{xxx}[\Omega^{(1)}_x]^2\over2[\Omega^{(0)}_{xx}]^3}~\!.
  \label{eqn:x2Determined}
  \end{eqnarray}
$y_1$ and $y_2$ are given by the analogous formulae. Inserting the corrections
to the chemical potentials determined in this way into the formulae for
$F^{(2)}$ and $F^{(3)}$ one finds that (again all functions are evaluated at
$x_0$ and $y_0$ or at $\tilde x_0$ and $\tilde y_0$) 
\begin{eqnarray}
F^{(2)}=\Omega^{(2)}
-{[\Omega^{(1)}_x]^2\over2~\!\Omega^{(0)}_{xx}}
-{[\Omega^{(1)}_y]^2\over2~\!\Omega^{(0)}_{yy}}~\!.\label{eqn:F(2)}
\end{eqnarray}
and (the formulae for $x_1$ and $y_1$ immediately imply that the first
four terms in the last line of the formula (\ref{eqn:F}) sum up to zero) that
\begin{eqnarray}
F^{(3)}=\Omega^{(3)}-{\Omega^{(2)}_x\Omega^{(1)}_x\over\Omega^{(0)}_{xx}}
-{\Omega^{(2)}_y\Omega^{(1)}_y\over\Omega^{(0)}_{yy}}
\phantom{aaaaaaaaaaaaaaaaaaaaaaaaaaaaaaaaaaaaaaaa}~\label{eqn:F(3)}\\
+~\!{\Omega^{(1)}_{xx}[\Omega^{(1)}_x]^2\over2~\![\Omega^{(0)}_{xx}]^2}
+{\Omega^{(1)}_{yy}[\Omega^{(1)}_y]^2\over2~\![\Omega^{(0)}_{yy}]^2}
+{\Omega^{(1)}_{xy}\Omega^{(1)}_x\Omega^{(1)}_y\over\Omega^{(0)}_{xx}\Omega^{(0)}_{yy}}
-{\Omega^{(0)}_{xxx}[\Omega^{(1)}_x]^3\over6~\![\Omega^{(0)}_{xx}]^3}
-{\Omega^{(0)}_{yyy}[\Omega^{(1)}_y]^3\over6~\![\Omega^{(0)}_{yy}]^3}~\!.
\nonumber
\end{eqnarray}
It will be seen that the extra terms in (\ref{eqn:F(2)}) precisely cancel the
contributions to $\Omega^{(2)}$ of those diagrams which do not contribute to
the expansion of the formula (\ref{eqn:EOmegaFormula}) for the ground state
energy density. The analogous cancellation of the extra terms in
(\ref{eqn:F(3)}) and the in $\Omega^{(3)}$ is demonstrated in Appendix B.

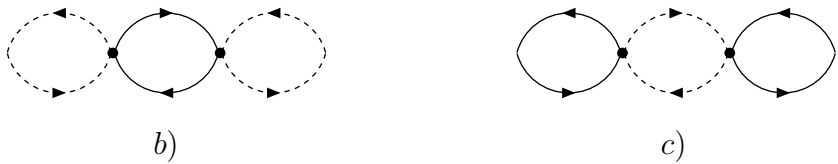
\begin{figure}[]
\begin{center}
\begin{picture}(300,40)(5,0)
\DashArrowArc(10,35)(20,195,345){2}
\DashArrowArc(10,25)(20,15,165){2}
\Vertex(30,30){2}
\ArrowArcn(50,35)(20,345,195)
\ArrowArcn(50,25)(20,165,15)
\Vertex(70,30){2}
\DashArrowArc(90,35)(20,195,345){2}
\DashArrowArc(90,25)(20,15,165){2}
\Text(50,-5)[]{$b)$}
\ArrowArc(200,35)(20,195,345)
\ArrowArc(200,25)(20,15,165)
\Vertex(220,30){2}
\DashArrowArcn(240,35)(20,345,195){2}
\DashArrowArcn(240,25)(20,165,15){2}
\Vertex(260,30){2}
\ArrowArc(280,35)(20,195,345)
\ArrowArc(280,25)(20,15,165)
\Text(240,-5)[]{$c)$}
\end{picture}
\end{center}
\caption{The order $C^2_0$ contributions $\Omega^{(2)b}$ and $\Omega^{(2)c}$.}
\label{fig:C0squareOther}
\end{figure}

\section{Computation of $F^{(2)}$}

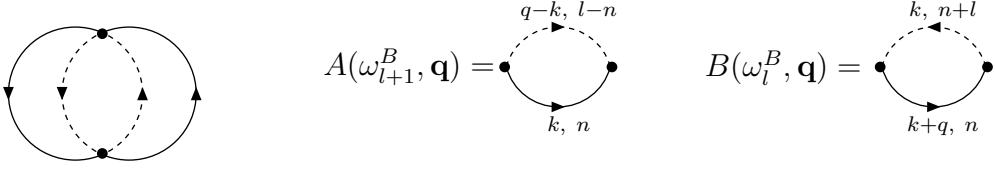
\begin{figure}[]
\begin{center}
\begin{picture}(370,40)(5,0)
\ArrowArc(30,20)(25,70,290)
\DashArrowArc(30,20)(25,290,70){2}
\DashArrowArc(50,20)(25,110,250){2}
\ArrowArc(50,20)(25,250,110)
\Vertex(40,-2.5){2}
\Vertex(40,42.5){2}
\Text(155,30)[]{$A(\omega_{l+1}^B,\mathbf{q})=$}
\Vertex(190,30){2}
\ArrowArc(210,35)(20,195,345)
\DashArrowArcn(210,25)(20,165,15){2}
\Vertex(230,30){2}
\Text(215,50)[]{$^{q-k,~l-n}$}
\Text(215,8)[]{$_{k,~n}$}
\Text(295,30)[]{$B(\omega_l^B,\mathbf{q})=$}
\Vertex(330,30){2}
\ArrowArc(350,35)(20,195,345)
\DashArrowArc(350,25)(20,15,165){2}
\Vertex(370,30){2}
\Text(355,50)[]{$^{k,~n+l}$}
\Text(355,8)[]{$_{k+q,~n}$}
\end{picture}
\end{center}
\caption{The order $C_0^2$ diagram contributing to the
  correction $\Omega^{(2)}$ and
  two ``elementary'' one-loop diagrams out of which the second
  order and the third order corrections with the $C_0$ couplings can be
  constructed. Solid and dashed lines denote propagators of fermions
  with the spin projections $+$ and $-$, respectively.}
\label{fig:ElementaryLoops}
\end{figure}

Diagrams contributing to $\Omega^{(2)}$ are shown in Figures
\ref{fig:C0squareOther} and \ref{fig:ElementaryLoops} (the left one).
It is straightforward to check that the contributions $\Omega^{(2)b}$
and $\Omega^{(2)c}$ of the ones of Figure \ref{fig:C0squareOther}
cancel against the last two terms in the formula (\ref{eqn:F(2)}).
Indeed, with the help of the summation rules collected in
Appendix A and taking into account that these contributions
are evaluated at $x_0$ and $y_0$ one easily obtains
($\Omega^{(2)c}$ is given by an analogous formula)
\begin{eqnarray}
  \Omega^{(2)b}={C_0^2V\over2}
       \left({N_-\over V}\right)^2
       \int\!{d^3\mathbf{p}\over(2\pi)^3}\left[{d\over da}~\!
         {1\over1+e^{\beta a}}\right]_{a=\varepsilon_{\mathbf{p}}-x_0}
       =-{1\over2}~\!C_0^2V\beta~\!(n_x-n_{xx})n_y^2~\!,
       \label{eqn:C0squared2b}
\end{eqnarray}
where the second form of $\Omega^{(2)b}$ is given 
in the notation introduced in Appendix B. With the
help of the formulae (\ref{eqn:Omega0Derivs}),
(\ref{eqn:Omega1Derivs}) it is immediately seen that it is canceled
by the second term of (\ref{eqn:F(2)}). Thus
\begin{eqnarray}
  \Omega^{(2)b}+\Omega^{(2)c}-{[\Omega^{(1)}_x]^2\over2~\!\Omega^{(0)}_{xx}}
-{[\Omega^{(1)}_y]^2\over2~\!\Omega^{(0)}_{yy}}=0~\!.\nonumber
\end{eqnarray}
Hence, $F^{(2)}=\Omega^{(2)a}$ evaluated at $x_0$ and $y_0$ (or at $\tilde x_0$
and $\tilde y_0$).

The integrals and sums corresponding to the left diagram of Figure
\ref{fig:ElementaryLoops} giving $\Omega^{(2)a}$ 
can be written in three different forms (corresponding to three different
routings of the internal momenta and frequencies) of which two can be
composed of two ``elementary'' blocks $A$ and $B$ shown in Figure
\ref{fig:ElementaryLoops} right:
\begin{eqnarray}
  \Omega^{(2)a}=-{1\over2}~\!C_0^2V~\!{1\over\beta}\sum_{l\in\mathbb{Z}}\!
  \int\!{d^3\mathbf{q}\over(2\pi)^3}~\![A(\omega_l^B,\mathbf{q})]^2
  =-{1\over2}~\!C_0^2V~\!{1\over\beta}\sum_{l\in\mathbb{Z}}\!
  \int\!{d^3\mathbf{q}\over(2\pi)^3}~\![B(\omega_l^B,\mathbf{q})]^2.\nonumber
\end{eqnarray}
where (here
$n_\pm(\mathbf{p})\equiv[1+\exp\{\beta(\varepsilon_{\mathbf{p}}-\mu^{(0)}_\pm)\}]^{-1}$,
$\mu^{(0)}_+\equiv x_0$, $\mu^{(0)}_-\equiv y_0$)
\begin{eqnarray}
  A(\omega_{l+1}^B,\mathbf{q})={1\over\beta}\sum_{n\in\mathbb{Z}}\!
  \int\!{d^3\mathbf{k}\over(2\pi)^3}~\!
  {1\over i\omega_n^F-(\varepsilon_{\mathbf{k}}-x_0)}~\!
  {1\over i\omega_{l-n}^F-(\varepsilon_{\mathbf{q}-\mathbf{k}}-y_0)}\nonumber\\
  =\int\!{d^3\mathbf{k}\over(2\pi)^3}~\!
  {n_+(\mathbf{k})+n_-(\mathbf{q}-\mathbf{k})-1\over
      i\omega_{l+1}^B-(\varepsilon_{\mathbf{k}}-x_0
    +\varepsilon_{\mathbf{q}-\mathbf{k}}-y_0)}~\!.\phantom{aaaaaaaaa}\label{eqn:Ablock}
\end{eqnarray}
and
\begin{eqnarray}
  B(\omega_l^B,\mathbf{q})={1\over\beta}\sum_{n\in\mathbb{Z}}\!
  \int\!{d^3\mathbf{k}\over(2\pi)^3}~\!
  {1\over i\omega_n^F-(\varepsilon_{\mathbf{k}+\mathbf{q}}-x_0)}~\!
  {1\over i\omega_{n+l}^F-(\varepsilon_{\mathbf{k}}-y_0)}
  \nonumber\\
  =\int\!{d^3\mathbf{k}\over(2\pi)^3}~\!
  {n_+(\mathbf{k}+\mathbf{q})-n_-(\mathbf{k})\over
    i\omega_l^B-(\varepsilon_{\mathbf{k}}-y_0-\varepsilon_{\mathbf{k}+\mathbf{q}}+x_0)}
  ~\!.\phantom{aaaaaaaaaa}~\label{eqn:Bblock}
\end{eqnarray}
(The contributions $\Omega^{(3)a}$ and $\Omega^{(3)b}$
of the left and right diagrams shown in Figure \ref{fig:C0cubeMercedes}
can be written analogously with $[A(\omega_l^B,\mathbf{q})]^3$ and
$[B(\omega_l^B,\mathbf{q})]^3$, respectively \cite{CHWO3,CHWO4}).

With the help of the sum rule (\ref{eqn:SSB1}) the sum over $l$ of two
$A$-blocks can be done and gives (the symbol $\int_{\mathbf{k}}$ stands for
the integral over the measure $d^3\mathbf{k}/(2\pi)^3$)
\begin{eqnarray}
  \int_{\mathbf{k}}\!\int_{\mathbf{p}}
  {[n_+(\mathbf{k})+n_-(\mathbf{q}-\mathbf{k})-1]
  [n_+(\mathbf{p})+n_-(\mathbf{q}-\mathbf{p})-1]\over
  \varepsilon_{\mathbf{k}}+\varepsilon_{\mathbf{q}-\mathbf{k}}
  -\varepsilon_{\mathbf{p}}-\varepsilon_{\mathbf{q}-\mathbf{p}}}
  \phantom{aaaaaaaaaaaaaaaaaa}\nonumber\\
  \times\left({1\over1-e^{\beta(\varepsilon_{\mathbf{k}}-x_0)}
    e^{\beta(\varepsilon_{\mathbf{q}-\mathbf{k}}-y_0)}}
  -{1\over1-e^{\beta(\varepsilon_{\mathbf{p}}-x_0)}
    e^{\beta(\varepsilon_{\mathbf{q}-\mathbf{p}}-y_0)}}\right).\nonumber
\end{eqnarray}
The identity
\begin{eqnarray}
  n_+(\mathbf{k})+n_-(\mathbf{q}-\mathbf{k})-1
  =n_+(\mathbf{k})~\!n_-(\mathbf{q}-\mathbf{k})\left[1-e^{\beta(\varepsilon_{\mathbf{k}}-x_0)}
    e^{\beta(\varepsilon_{\mathbf{q}-\mathbf{k}}-y_0)}\right].\label{eqn:Identity1}
\end{eqnarray}
and the fact that the two terms in the bracket above
gives equal contributions allows then to write
\begin{eqnarray}
  \Omega^{(2)a}=C_0^2V\!\int_{\mathbf{q}}\!\int_{\mathbf{p}}\!\int_{\mathbf{k}}\!
  {n_+(\mathbf{k})~\!n_-(\mathbf{q}-\mathbf{k})[1-
  n_+(\mathbf{p})-n_-(\mathbf{q}-\mathbf{p})]\over
  \varepsilon_{\mathbf{k}}+\varepsilon_{\mathbf{q}-\mathbf{k}}
  -\varepsilon_{\mathbf{p}}-\varepsilon_{\mathbf{q}-\mathbf{p}}}~\!.
  \label{eqn:Omega2aAA}
\end{eqnarray}
It is interesting to notice that because the integral of the
quartic product
$n_+(\mathbf{k})~\!n_-(\mathbf{q}-\mathbf{k})n_+(\mathbf{p})~\!n_-(\mathbf{q}-\mathbf{p})$
vanishes (the numerator is even with respect to the interchange
$\mathbf{k}\leftrightarrow\mathbf{p}$ while the denominator is odd),
the expression for $\Omega^{(2)a}$ can be written (after the change
$\mathbf{p}=-\mathbf{u}+\mathbf{s}$, 
$\mathbf{k}=-\mathbf{t}+\mathbf{s}$, $\mathbf{q}=2\mathbf{s}$
of the integration variables) in the form completely analogous to the
expression giving $E_\Omega/V$ (see \cite{CHWO1}), the only modification being the
change in the prefactor
and the replacement of $\theta(k-|\mathbf{v}|)$ and $\theta(|\mathbf{v}|-k)$
by $n(|\mathbf{v}|)$ and $1-n(|\mathbf{v}|)$, respectively. (Curiously
enough, we have found that this simple analogy does not work for
the diagrams of Figure \ref{fig:C0cubeMercedes}).

It is straightforward to see that the expression (\ref{eqn:Omega2aAA})
is divergent, the divergence arising from the unity in the square
bracket in the numerator. In the variables $\mathbf{s}$, $\mathbf{t}$ and
$\mathbf{u}$ the integral over $\mathbf{u}$ is the one evaluated with the
cutoff $\Lambda$ in \cite{CHWO1} and using this result and changing once
more the variables to  $\mathbf{k}=\mathbf{t}-\mathbf{s}$,
$\mathbf{p}=\mathbf{t}+\mathbf{s}$, after adding the contribution
$\Omega^{(1)}$ and expressing $C_0$ in terms of the scattering lengths $a_0$
\begin{eqnarray}
  C_0(\Lambda)={4\pi\hbar^2\over m_f}~\!a_0\left(1
  +{2\over\pi}~\!a_0\Lambda+\dots\right), \label{eqn:C0Determined}
\end{eqnarray}
\cite{WeDrSch,CHWO3,CHWO4}, one arrives at the finite (to the second order)
result
\begin{eqnarray}
  \Omega^{(1)}+\Omega^{(2)a}={4\pi\hbar^2\over m_f}~\!a_0\left(1
  +{2\over\pi}~\!\Lambda a_0+\dots\right)V~\!
  {N_-\over V}~\!{N_+\over V}
  \phantom{aaaaaaaaaaaaaaaaaaaaaaaaaaa}\nonumber\\
  -~\!{\Lambda\over2\pi^2}\left({4\pi\hbar^2\over m_f}~\!a_0\right)^2
  {m_f\over\hbar^2}~\!V
  ~\!{N_-\over V}~\!{N_+\over V}+\Omega^{(2)a}_{\rm finite}
 ={4\pi\hbar^2\over m_f}~\!a_0~\!V~\!
  {N_-\over V}~\!{N_+\over V}+\Omega^{(2)a}_{\rm finite}~\!.\nonumber
\end{eqnarray}
The finite part of $\Omega^{(2)a}$,
\begin{eqnarray}
  \Omega^{(2)a}_{\rm fin}=-C_0^2V\!\int_{\mathbf{q}}\int_{\mathbf{k}}n_+(\mathbf{k})~\!
  n_-(\mathbf{q}-\mathbf{k})\int_{\mathbf{p}}
  {n_+(\mathbf{p})+n_-(\mathbf{q}-\mathbf{p})\over
    \varepsilon_{\mathbf{k}}+\varepsilon_{\mathbf{q}-\mathbf{k}}
    -\varepsilon_{\mathbf{p}}-\varepsilon_{\mathbf{q}-\mathbf{p}}}~\!,\nonumber
\end{eqnarray}
upon setting first  $\mathbf{k}=\mathbf{k}_1$,
$\mathbf{q}-\mathbf{k}=\mathbf{k}_2$ and then replacing
in the term with $n_-(\mathbf{k}_1+\mathbf{k}_2-\mathbf{p})$
the variable $\mathbf{k}_1+\mathbf{k}_2-\mathbf{p}$ by $\mathbf{p}^\prime$ (upon which
$\varepsilon_{\mathbf{k}_1+\mathbf{k}_2-\mathbf{p}}\rightarrow\varepsilon_{\mathbf{p}^\prime}$
but at the same time
$\varepsilon_{\mathbf{p}}\rightarrow\varepsilon_{\mathbf{k}_1+\mathbf{k}_2-\mathbf{p}^\prime}$)
can be cast in the convenient symmetric form
\begin{eqnarray}
  \Omega^{(2)a}_{\rm fin}=F^{(2)}
  =-C_0^2V\!\int_{\mathbf{k}_1}\int_{\mathbf{k}_2}n_+(\mathbf{k}_1)~\!
  n_-(\mathbf{k}_2)\int_{\mathbf{p}}
  {n_+(\mathbf{p})+n_-(\mathbf{p})\over
    \varepsilon_{\mathbf{k}_1}+\varepsilon_{\mathbf{k}_2}
    -\varepsilon_{\mathbf{p}}-\varepsilon_{\mathbf{k}_1+\mathbf{k}_2-\mathbf{p}}}~\!.
  \label{eqn:Omega(2)a}
\end{eqnarray}
The expression (\ref{eqn:Omega(2)a}) is very similar\footnote{Recall
  the standard rule
  $\sum_{\mathbf{k}}\rightarrow V\int d^3\mathbf{k}/(2\pi)^3\equiv V\int_{\mathbf{k}}$
  for passing from the box normalization used in \cite{Pathria,DUMacDO}
  to the continuum one used here.}
to the formula (5) used in \cite{DUMacDO} as the second order contribution to
the system's internal energy density $u$, except that the latter has an extra
factor of 2. The foundation of the formula for $f=u-Ts$ (which, apart from
this factor of 2, is equivalent to our one) used in \cite{DUMacDO} is,
however, somewhat unclear: to obtain their second order correction to the
energy density $u$ these authors took the expression (15) given
in Section 11.4 of \cite{Pathria} which is obtained by simply using the
finite temperature distributions in place of the zero temperature ones
in the ordinary second order correction to the ground state energy
of the system and have taken the entropy density $s$ as given
by the zeroth order textbook formula.
In contrast our expression (\ref{eqn:Omega(2)a}) results from a
systematic, well founded expansion and the coefficient in (\ref{eqn:Omega(2)a})
is unambiguously fixed by the cancellation of the divergence.

After integrating over the cosine of the angle between $\mathbf{p}$ and
$\mathbf{k}_1+\mathbf{k}_2$ one can write the resulting expression in the
form
\begin{eqnarray}
  F^{(2)}=-V~\!{C_0^2m_f\over(2\pi)^2\hbar^2}
  \!\int_{\mathbf{k}_1}\int_{\mathbf{k}_2}
   {n_+(\mathbf{k}_1)~\!n_-(\mathbf{k}_2)\over|\mathbf{k}_1+\mathbf{k}_2|}
   \int_0^\infty\!dp~\!p\left[n_+(\mathbf{p})+n_-(\mathbf{p})\right]
   \phantom{aaaaaaaaa}\nonumber\\
   \times\ln\!\left|{(p-\Delta_+)(p-\Delta_-)\over(p+\Delta_+)(p+\Delta_-)}\right|
  ,\label{eqn:TripleIntegration}
\end{eqnarray}
in which
\begin{eqnarray}
  \Delta_\pm\equiv{1\over2}~\!|\mathbf{k}_1+\mathbf{k}_2|\pm
  {1\over2}~\!|\mathbf{k}_1-\mathbf{k}_2|~\!. \nonumber
\end{eqnarray}
It is clear that the singularity at $|\mathbf{k}_1+\mathbf{k}_2|=0$
in (\ref{eqn:TripleIntegration}) is spurious:
if $\mathbf{k}_1+\mathbf{k}_2=\mathbf{0}$ then $\Delta_-=-\Delta_+$
and the innermost integral vanishes.

\section{Numerical evaluation}

The most difficult part of the computation is the accurate and efficient
numerical evaluation of the multiple integrals in the expression
(\ref{eqn:TripleIntegration}). Rescaling the momentum integration
variables $\mathbf{k}_1=k_{\rm F}\mathbf{v}_1$, etc. and inserting
$C_0=(4\pi\hbar/m_f)a_0$ one can write the second order contribution
to the right hand side of (\ref{eqn:FtoFirstOrderDimless}) as
\begin{eqnarray}
  {6\pi^2\over k_{\rm F}^3}~\!{F^{(2)}\over\varepsilon_{\rm F}V}
  =-(k_{\rm F}a_0)^2~\!{6\over\pi^2}\int_0^\infty\!dv_1~\!v_1^2~\!n(v_1,\nu_+,t)\!
  \int_0^\infty\!dv_2~\!v_2^2~\!n(v_2,\nu_-,t)\nonumber\\
  \times\sum_{\sigma=\pm}\sum_{\sigma^\prime=\pm}\!\int_{-1}^1\!d\xi~\!
  {I(\Delta_\sigma,\nu_{\sigma^\prime},t)\over\sqrt{v_1^2+v_2^2+2\xi v_1v_2}}~\!,
  \phantom{aaa}\label{eqn:f2}
  \end{eqnarray}
where $\nu_\pm=\mu^{(0)}_\pm/k_{\rm B}T\equiv\delta_\pm/t$
($\delta_\pm=\mu^{(0)}_\pm/\varepsilon_{\rm F}$),
\begin{eqnarray}
  n(v,\nu,t)=\left[1+\exp\!\left({v^2\over t}-\nu\right)\right]^{-1}, \nonumber
\end{eqnarray}
\begin{eqnarray}
  I(\Delta,\nu,t)=\int_0^\infty\!du~\!u~\!n(u,\nu,t)~\!
  \ln\!\left|{u-\Delta\over u+\Delta}\right|.  \nonumber
\end{eqnarray}
and
\begin{eqnarray}
  \Delta_\pm(v_1,v_2,\xi)={1\over2}\sqrt{v_1^2+v_2^2+2\xi v_1v_2}
  \pm{1\over2}\sqrt{v_1^2+v_2^2-2\xi v_1v_2}~\!.\nonumber
\end{eqnarray}
The trick allowing to realize the numerical computation is to make first,
for fixed values of $t$ (temperature) and $P$ (the system's polarization)
which together, through (\ref{eqn:nuDetermination}) determine $\nu_+$ and
$\nu_-$, an interpolation of the functions $I(|\Delta|,\nu_+,t)$ and
$I(|\Delta|,\nu_-,t)$ (because, obviously,
$I(-|\Delta|,\nu_\pm,t)=-I(|\Delta|,\nu_\pm,t)$) in the variable
$w=1/(1+|\Delta|)$
(to interpolate on the
compact interval $[0,1]$) and then performing numerically the integrations
over $v_1$, $v_2$ and $\xi$ using these interpolations.
In the actual code written in the {\it Python} programming language
the functions $I(|\Delta|,\nu_\pm,t)$ are evaluated with the help of
the adaptive integration routine (scipy.integrate.quad;
the integration domain is split ted into three subdomains to accurately
handle the logarithmic singularity - in the relevant regions
near $w =1/(1+\Delta)\equiv w_0$ we substitute $r^3 = |w - w_0|$
so that the integrand behaves like $r^2\ln(r)$ and can be treated
using the quadrature methods -
and its sharp
falloff, especially for small temperatures $t$, near $u^2=t\nu$ of the
distribution $n(u,\nu,t)$) and then interpolated using the
cubic spline interpolation routines of  {\it Python}.
The remaining triple integral over $v_1$, $v_2$ and $\xi$ are
again performed with the help of the Clenshaw-Curtis quadrature in the
variables $w_{1,2}=1/(1+v_{1,2})$ (again to have
a compact integration domain and again splitting it into subdomains
to better handle the regions $v_1^2\approx t\nu_+$ and 
$v_2^2\approx t\nu_-$); the spurious singularity at
$|\mathbf{v}_1+\mathbf{v}_2|=0$ is taken care of
by simply taking somewhat different numbers for the $v_1$ and $v_2$ grids.

To check the correctness of the code we have first compared its results
for $t\rightarrow0$ (replacing the distributions $n(u,\nu,t)$ by the
Heaviside theta functions) with the second order correction $E^{(2)}_\Omega$
to the system's ground-state energy which as a function of $P$ is known
analytically \cite{KANNO,PECABO} (the function $J_K(x,y)$ is given e.g. by
the formula (4) in \cite{CHWO4}):
\begin{eqnarray}
  {6\pi^2\over k_{\rm F}^3}~\!{E^{(2)}_\Omega\over\varepsilon_{\rm F}V}
  =(k_{\rm F}a_0)^2~\!{6\over5\pi^2}~\!J_K((1+P)^{1/3},~\!(1-P)^{1/3})~\!.\nonumber
\end{eqnarray}
At $P=0$ (equal densities of spin up and spin down fermions) $J_K=4(11-\ln4)/21$
and the right hand side of the above formula (setting in all these comparisons
$k_{\rm F}a_0=1$) equals $0.22264482$ while the {\it Python} code for the right
hand side of (\ref{eqn:f2}) gives the value $0.22264522$. For $P=0.5$ the code
gives $0.17184256$ to be compared with $0.17184207$ while at $P=0.9$ the numbers
to be compared are $0.046470057$ and $0.046470077$ 
(at $P=1$ both are zero reflecting the impossibility of the
$s$-wave interactions of two fermions in the same spin state).
For nonzero temperatures the results obtained using the Clenshaw-Curtis
quadrature have been compared with the ones obtained using the more accurate
(but more time consuming) adaptive integration routine. The comparison shows
that the relative uncertainty $\Delta_F$ (the difference of the results of
the two methods divided by their mean) is typically of order $10^{-5}$,
varying rather irregularity with $P$ and increasing somewhat with $t$;
in our further estimates we set $\Delta_F=10^{-5}$ for $t\simlt0.1$,
$\Delta_F=1.5\times10^{-5}$ for $0.1< t\leq0.2$ and 
$\Delta_F=2\times10^{-5}$ for $0.2<t$.
While this accuracy superficially looks quite satisfactory, it is, nevertheless,
barely sufficient: for values of the parameters ($t$ and/or $k_{\rm F}a_0$)
at which spontaneous ordering appears there is a very delicate cancellation
between different contributions to $F$ and the (relative) error of order
$10^{-5}$ in $F^{(2)}$ can, and in some cases indeed does, lead do the appearances
of very shallow fake minimum near $P=0$.

\section{Results}

\begin{figure}
\centerline{\hbox{
\psfig{figure=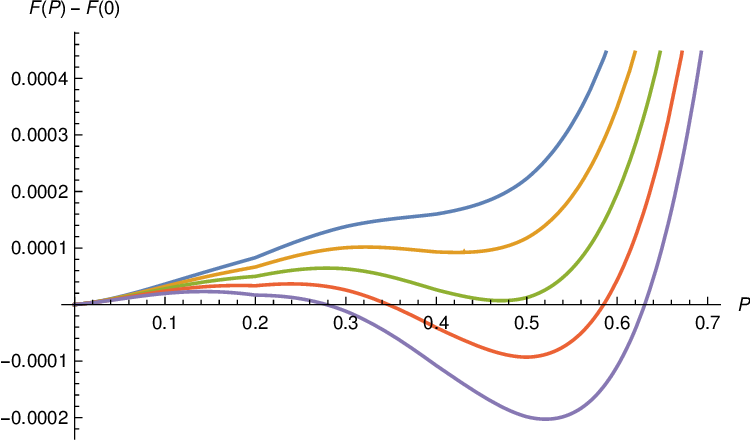,width=8.cm,height=5.0cm} 
\psfig{figure=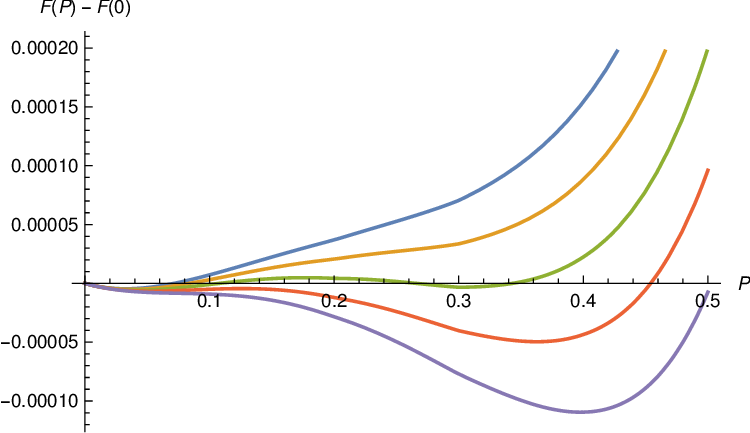,width=8.cm,height=5.0cm} 
}}
\caption{Plots of the differences $F(P)-F(0)$ in units of
  $(k_{\rm F}^3/6\pi^2)(\hbar^2k_{\rm F}^2/2m_f)$ of the system of spin $1/2$
  fermions as a function of the order parameter $P$ for two representative
  values of the temperature $t\equiv T/T_{\rm F}$ as obtained in the second
  order of the perturbative expansion.
  Left: $t=0.1$; the successive curves (from below) correspond to
  $k_{\rm F}a_0=1.0718$ (the lowest, blue, line), $k_{\rm F}a_0=1.0723$
  (yellow), $k_{\rm F}a_0=1.0728$ (green), 
  $1.0733$ (red) and $1.0738$ (the highest, blue, line).
  Right: $t=0.15$; the successive curves (from below) correspond to
  $k_{\rm F}a_0=1.0978$ (the lowest, blue, line), $k_{\rm F}a_0=1.983$
  (yellow), $k_{\rm F}a_0=1.0988$ (green), 
  $1.0993$ (red) and $1.0998$ (the highest, blue, line).}
\label{fig:F2forFixedT}
\end{figure}

For a fixed value of the temperature, the system's free energy $F$ as a
function of the polarization (and of the parameter $k_{\rm F}a_0$ in which,
in the approximation to which our analysis is restricted, it is a polynomial
of the second order) can be efficiently obtained by evaluating numerically
the integrals in (\ref{eqn:f2}) for several values of $P$ and constructing
the cubic spline interpolation. The resulting free energy differences,
$F(P)-F(0)$, are plotted in Figure \ref{fig:F2forFixedT} as functions of the
polarization $P$ for two temperatures: $t=0.1$ and $0.15$ and several values
of $k_{\rm F}a_0$ (obtained by constructing the interpolation based on
11 points in $P$ only).

In view of the mentioned uncertainty in the computation of $F^{(2)}$
the critical value of $k_{\rm F}a_0$ and the value of the polarization
$P$ at the transition must be determined by requiring that the value
of $F$ at a minimum developing away from $P=0$ differs from the one at
$P=0$ at least by $\Delta_F F^{(2)}(0)$. In this way one can properly
handle the mentioned fake minima close to $P=0$ one of which can be
observed in the right panel of Fig. \ref{fig:F2forFixedT} (for $t=0$ that
such a minimum is indeed produced by the inaccuracies of the numerical
code can be substantiated by comparing with the analytically known
dependence of the ground-state energy on $P$). The actual procedure which
has been adopted to determine the polarization and its uncertainty is
as follows. For a fixed value of the parameter $k_{\rm F}a_0$ (which is
successively increased from 0 in steps $\Delta(k_{\rm F}a_0)=0.001$) the
values of $F$ on a preliminary grid of $P$-values $P_n=n\Delta P$
with $n=0,\dots,n_{\rm max}=32$
are obtained. If the minimal value of $F$ occurs for $n_{\rm min}=n_{\rm max}$
the polarization is taken as the maximal ($P=1$); if $n_{\rm min}=0$ or
$|F(P_{n_{\rm min}})-F(0)|\leq\Delta_FF^{(2)}(0)$ the polarization is taken as
vanishing ($P=0$). If $n_{\rm min}\neq0,n_{\rm max}$ and 
$|F(P_{n_{\rm min}})-F(0)|>\Delta_FF^{(2)}(0)$ the polarization is taken as
truly nonvanishing. If it is nonvanishing for the first time (as far as
the increasing values of $k_{\rm F}a_0$ are concerned), one determines
$n_{\rm down}$ and $n_{\rm up}$ such that
$|F(P_{n_{\rm min}})-F(P_{n_{\rm down/up}})|<2\Delta_FF^{(2)}(0)$ (of course,
$n_{\rm down}=0$ and/or $n_{\rm up}=n_{\rm max}$ if these criteria cannot not be
fulfilled for intermediate values of $n$) and finds the values of
$F$ on a finer grid of $P$-values with $|P_{j+1}-P_j|=0.0001$ and
$P_{n_{\rm down}}\leq P_j\leq P_{n_{\rm up}}$. If the minimum of $F$ found on
the finer grid occurs for $P_{j_{\rm min}}<0.02$,
it is assumed that it is a numerical artifact and the polarization is
taken as vanishing. In the opposite case the polarization is taken
to be nonvanishing and that value of $k_{\rm F}a_0$ is recorded
as the critical one (for the considered temperature). In this 
case on the finer grid one seeks a range $(P_{j_{\rm down}},~\!P_{j_{\rm up}})$
of $P$ around $P_{j_{\rm min}}$ in which
$|F(P_{j_{\rm min}})-F(P_j)|>\Delta_F F^{(2)}(0)$
for $P_{j_{\rm down}}\leq P_j\leq P_{j_{\rm up}}$; if such a range cannot be
found the transition is classified as continuous (the polarization at the
considered temperature 
is assumed to increase continuously from zero as $k_{\rm F}a_0$ is increased)
while if a nontrivial range is obtained, the transition is classified as
first order and
$P_{j_{\rm min}}-P_{j_{\rm down}}$ and $P_{j_{\rm up}}-P_{j_{\rm min}}$
are taken as the uncertainties of the determination of the
polarization right at the transition.
For values of $k_{\rm F}a_0$ higher than the critical one (determined as
described above for the considered temperature) $F$ is evaluated on a
finer grid of points $P_j$ with $P_{n_{\rm min}-1}\leq P_j\leq P_{n_{\rm min}+1}$
and $P_{j+1}-P_j=0.001$ and the corresponding polarization is
determined as the position of the minimum of $F$ on this finer grid.
In this way one finds
that $(k_{\rm F}a_0)_{\rm cr}=1.05409$ at $t=0$ (which perfectly agrees
with the known value obtained by computing the system's ground-state
energy \cite{CHWO3} and with \cite{DUMacDO}), 
$(k_{\rm F}a_0)_{\rm cr}=1.05858$ at $t=0.05$,
$(k_{\rm F}a_0)_{\rm cr}=1.07282$ at $t=0.1$,
$(k_{\rm F}a_0)_{\rm cr}=1.09881$ at $t=0.15$ and
$(k_{\rm F}a_0)_{\rm cr}=1.13845$ at $t=0.2$.
The corresponding values of the polarization right at the transition
point are $P_{\rm cr}=0.575_{-0.019}^{+0.017}$ (again in agreement with the
value found in \cite{CHWO3}) $0.558_{-0.017}^{+0.017}$,
$0.477_{-0.021}^{+0.019}$, $0.325_{-0.048}^{+0.035}$ and $0.197_{-0.096}^{+0.045}$.

\begin{figure}
\centerline{\hbox{
\centering
\includegraphics[width = 0.45 \textwidth]{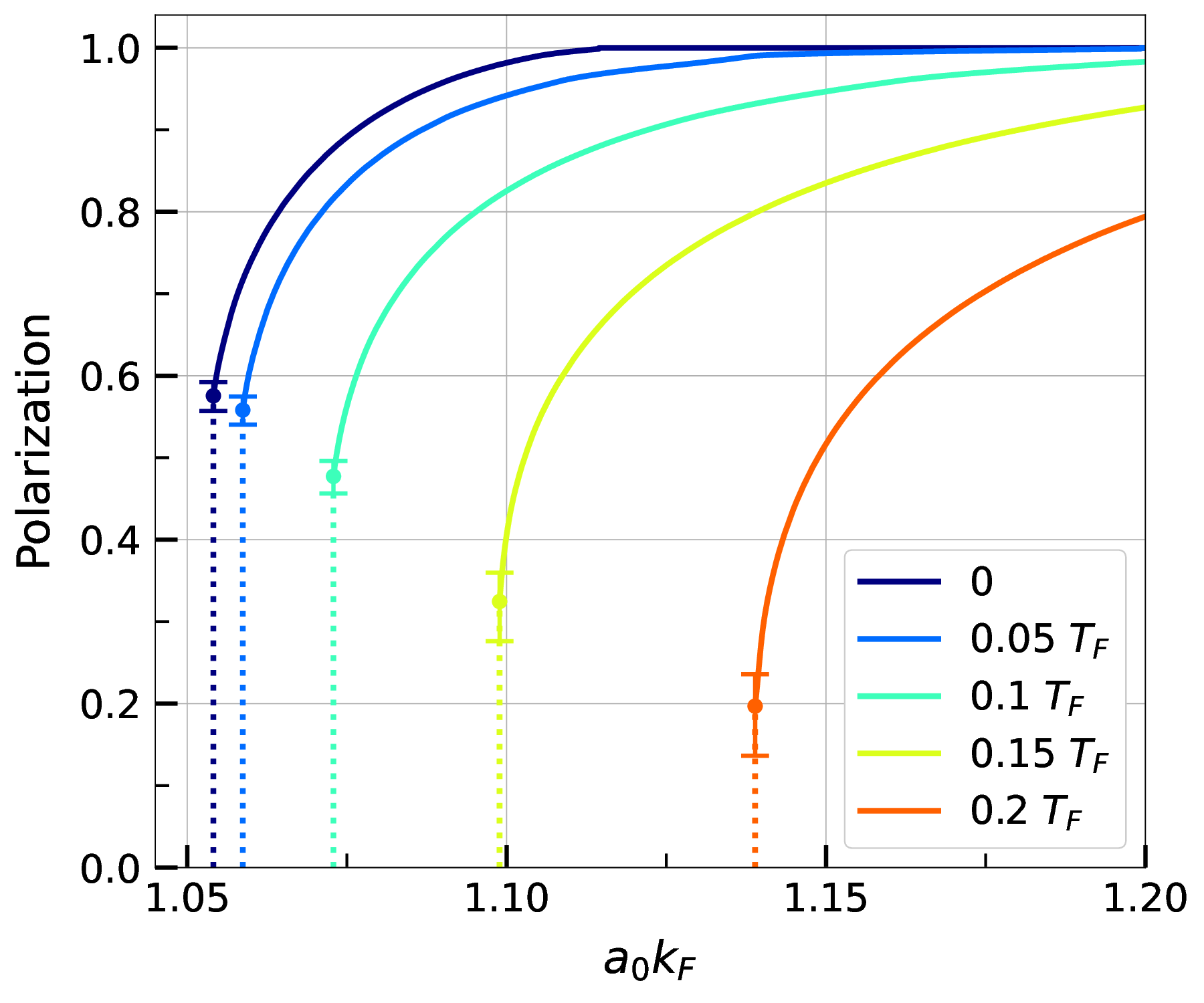}
\includegraphics[width = 0.45 \textwidth]{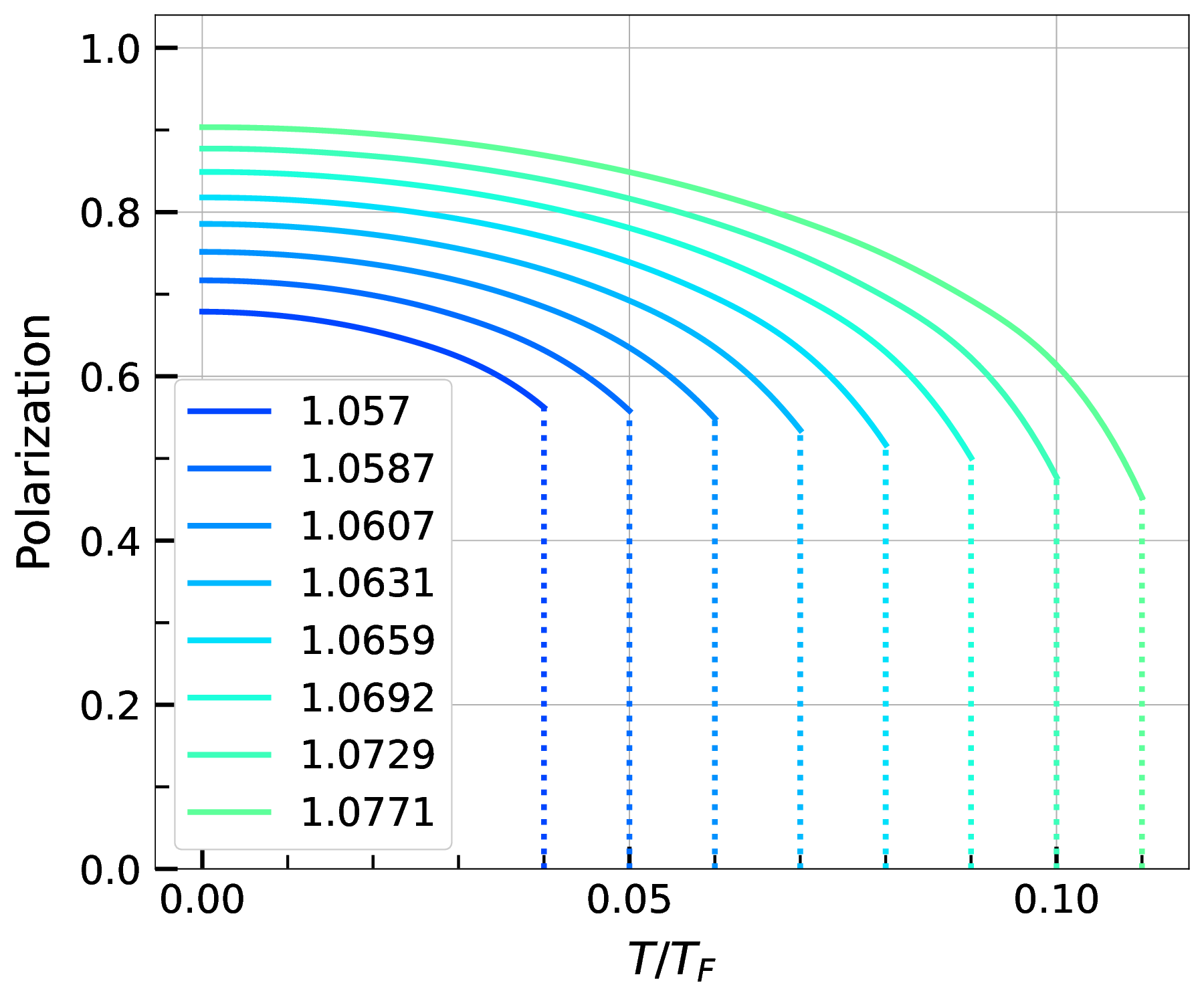}
}}
\caption{Polarization $P=(N_+-N_-)/N$ of the system of spin $1/2$
  fermions with a short range repulsive interaction obtained from
  the free energy $F$ computed up to the second order of the
  perturbative expansion. In the  left panel as a function of the
  ``gas parameter'' $k_{\rm F}a_0$ for several values of the temperature
  (counting from the left): $t\equiv T/T_F=0$, $0.1$, $0.15$ and
  $0.2$ ($T_{\rm F}\equiv \varepsilon_{\rm F}/k_{\rm B}$). Marked are also
  uncertainties of the value of $P$ right at the
  transition points. In the right panel as a function of the temperature
for several fixed values of $k_{\rm F}a_0$.}
\label{fig:Pol1}
\end{figure}

The dependence of the polarization  as a function of the ``gas parameter''
$k_{\rm F}a_0$ is shown, for a few values of the temperature $t$, in the left
panel of Figure \ref{fig:Pol1}. This is essentially the same plot as the
one presented in \cite{DUMacDO} (the agreement with the critical
values of the gas parameters at successive temperatures that can be read
off from the plot there seems to be quite good)
except that in Figure \ref{fig:Pol1} marked are also the uncertainties
in the determination (following from the procedure just described)
of the polarization right at the transition.

\begin{figure}
\centerline{\hbox{
\centering
\includegraphics[width = 0.45 \textwidth]{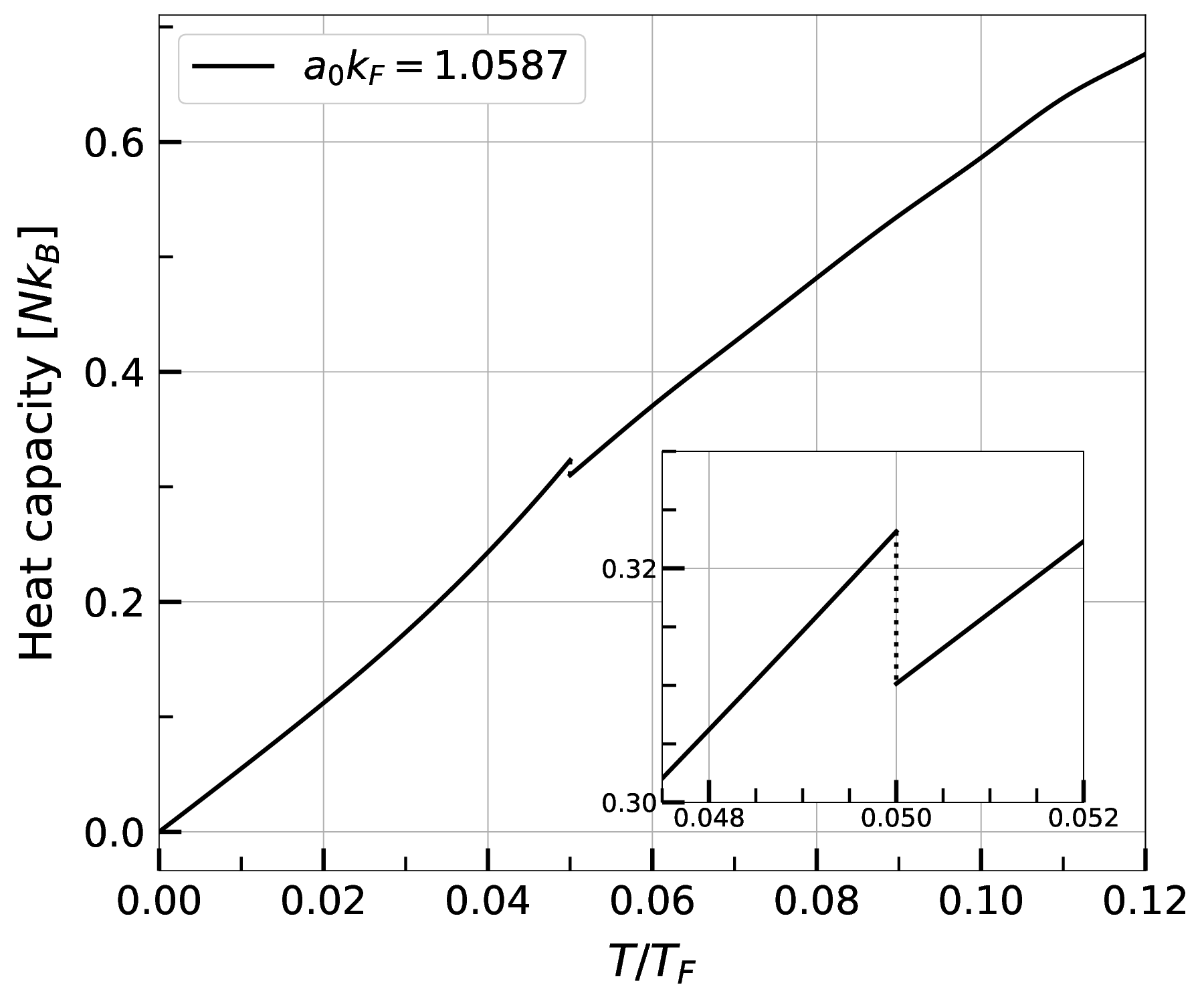}
\includegraphics[width = 0.45 \textwidth]{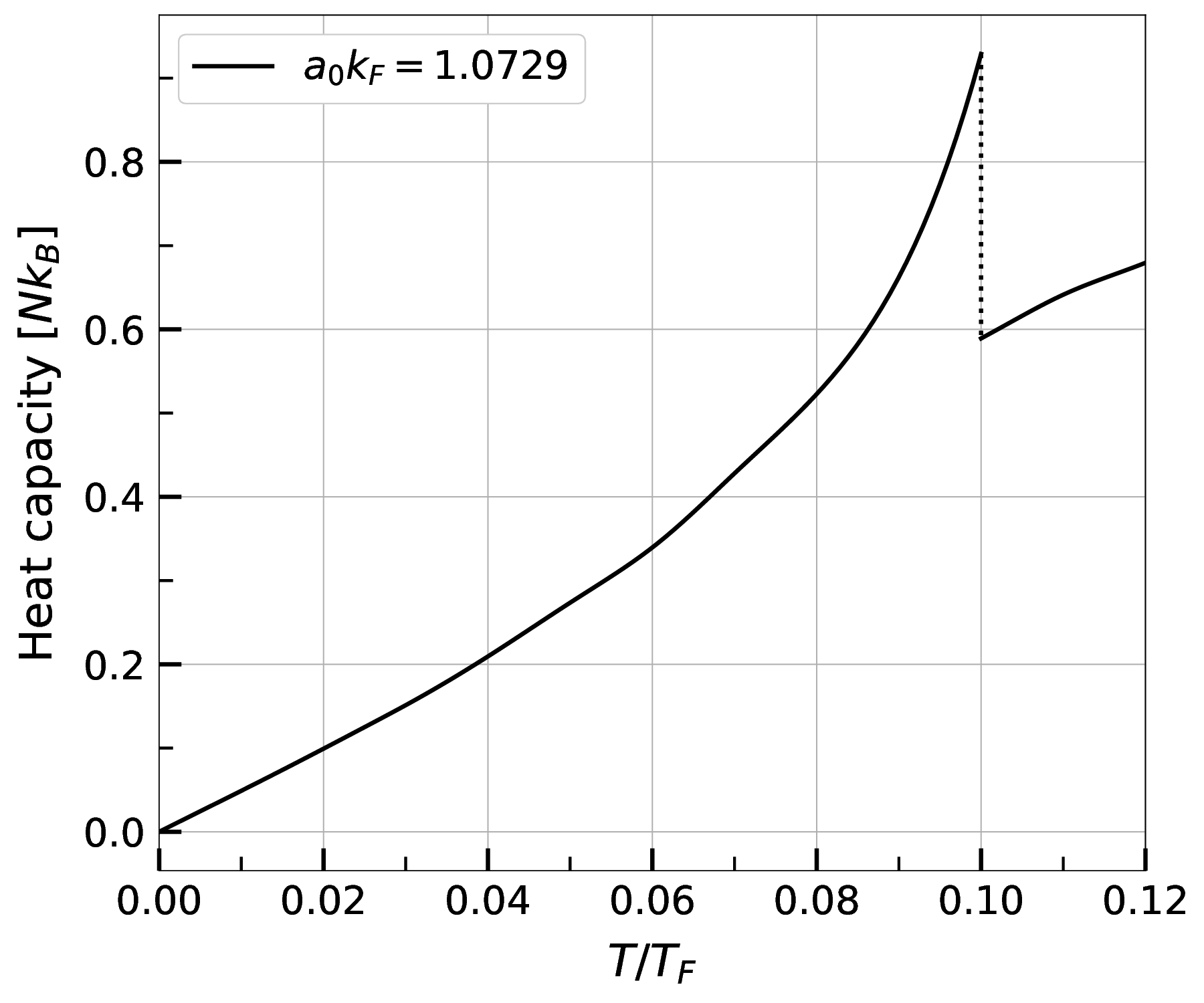}
}}
\caption{Heat capacity (in units of $Nk_{\rm B}$) of the system of spin $1/2$
  fermions with a short range repulsive interaction as a function
  of the temperature for two different fixed values of the
  parameter $k_{\rm F}a_0$ obtained from the free energy $F$
  computed up to the second order.}
\label{fig:HeatCapacity}
\end{figure}

Owing to the efficiency of our numerical code (stemming basically from
the trick with the interpolations) the procedure of finding the
polarization of the system described above can be applied also at
fixed values of $k_{\rm F}a_0$ (replacing the  grid in
$k_{\rm F}a_0$ by a one in $t$). The resulting polarization of the
system as a function of the temperature for
several fixed values of the gas parameter  is shown
in the right panel of Figure \ref{fig:Pol1}.
Knowing the polarization as a function of the
other parameters it is possible to
construct the free energy $F(T,V,N)\equiv F(T,V,N,P(T,N/V))$ for several
values of  $k_{\rm F}a_0$ and to determine also other thermodynamic
characteristics of the system. For example, using the grid in $t$
the second derivative of the free energy $F(T,V,N))$ with respect to the
temperature
can in principle be obtained yielding the system's heat capacity. The
result of such an exercise is shown in Figure \ref{fig:HeatCapacity}
for two values of the ``gas parameter''. It is shows that
the discontinuity of
the heat capacity at the transition point 
grows with the value of $k_{\rm F}a_0$ (i.e. also with the increasing
temperature, if $k_{\rm F}a_0$ is varied). However for higher
values of $k_{\rm F}a_0$ the numerical inaccuracies do not allow for
a reliable computation. Indeed, as the transition at higher
temperatures becomes continuous
a divergence of the heat capacity probably starts to build up making
the numerical computation of the second derivative of the free energy 
unstable for $t\simgt0.12$. Similarly, it is in principle possible
to determine the system's polarization taking into account an
infinitesimally weak external magnetic field (this as explained influences
only the determination of the zero-th order chemical potentials
$\tilde\mu^{(0)}_\pm$ from the conditions (\ref{eqn:mu0withH}))
and to compute the system's magnetic susceptibility $\chi_T$ by constructing
the derivative of the polarization with respect to ${\cal H}$.
While such a computation seems to indicate that at least at
low temperatures, at which
the transition is (in the approximation to which our computation
is restricted) is first order, the susceptibility also has 
a finite discontinuity at the transition point, it is not sufficiently
stable numerically to yield reliable values of $\chi_T$ and it is
probably more practical to obtain it by computing the
(connected) two point correlation function from the
formula $\chi_T=(\beta/V)\tilde G^{(2)}_{\rm con}(\mathbf{0})$.
We do not attempt this here.

\section{Conclusions}

We have developed a systematic perturbative expansion of the grand
thermodynamic potential $\Omega$ and of the free energy $F$ of the system
of (nonrelativistic) interacting spin 1/2 fermions. We have applied this
expansion within the effective field theory in which the underlying
repulsive spin-independent binary interaction of fermions is replaced
by an infinite number of contact interaction terms and which allows to
directly express computed quantities in terms of the scattering lengths
and effective radii which characterize the underlying interaction potential.
We have shown (up to the third order but the result seems to be valid in
general) that to the expansion of the free energy effectively contribute
only those Feynman diagrams which give nonvanishing contributions to the
ground-state energy of the system evaluated at zeroth order chemical
potentials (associated with spin up and spin down fermions).

Our numerical analysis has been restricted here to the first nontrivial
order of the perturbative expansion (i.e. the first one going beyond the
textbook mean field approximation) in which the results are still
universal, i.e. on the form of the underlying interaction depend only
through the $s$-wave scattering length $a_0$ (in the next order the
results start to depend also on the $p$-wave scattering length $a_1$ and
the effective radius $r_0$). We have devised a method for efficient
numerical evaluation of the requisite nested integrals and used it to
compute the system's polarization and its value right at the transition
point paying attention to the uncertainty of the determination of the
latter quantity which is crucial is assessing the character of the
transition. For low temperatures, $T\simlt0.1~\!T_{\rm F}$, we have also
managed to determine the system's heat capacity encountering, however,
some problems with the accuracy of numerical evaluation of the
derivatives of the free energy which seem to prevent obtaining (at
least without substantial improvements in the method)
reliable values of the heat capacity for higher temperatures as well
as determining the system's magnetic susceptibility.

Of course, since the perturbative computation of the system's
ground-state energy agrees with the results obtained (for specific
forms of the underlying interaction) using the Quantum Monte Carlo
approach only for $k_{\rm F}a_0\simlt0.5$, the results presented here
cannot be taken very seriously. Moreover it is now known that
already the inclusion of the third order corrections to the system's
ground-state energy (free energy at zero temperature) significantly weaken
the first order character of the transition (at zero temperature) to
the ordered state. For these reasons our effort summarized here
should be treated rather as a preliminary step taken towards
extending the computation to a higher order and towards a possible
implementation of a resummation of some class of the
contributions to the free energy in the spirit of the
approach of \cite{He}. Such a resummation can probably also
allow to overcome the limitation,  inherent in the effective
field theory approach, to sufficiently small temperatures only:
as this approach relies on the clean separation of the
 scales ($R\ll k_{\rm F}^{-1}$, where $R$ is the characteristic
length of the underlying interaction) it cannot
be applied, at least if restricted to a finite order
of the perturbative expansion,
when $k_{\rm B}T$ becomes comparable with
the energy scale set by $\varepsilon_{\rm F}$.
We plan to return to these issues in the forthcoming paper.

\vskip0.2cm

\setcounter{equation}{0}
\renewcommand{\theequation}{A.\arabic{equation}}

\section*{Appendix A}
The following summation formulae hold \cite{FetWal} (the limit
$\eta\rightarrow0^+$ is implicit):
\begin{eqnarray}
  &&{1\over\beta}\sum_{n\in\mathbb{Z}}{e^{i\eta\omega^F_n}\over i\omega^F_n-x}
  ={1\over1+e^{\beta x}}~\!,\phantom{aaa}\omega^F_n\equiv{\pi\over\beta}~\!
  (2n+1)~\!,
  \label{eqn:S1F}\\
  &&{1\over\beta}\sum_{n\in\mathbb{Z}}{e^{i\eta\omega^B_n}\over i\omega^B_n-x}
  ={1\over1-e^{\beta x}}~\!,\phantom{aaa}\omega^B_n\equiv{2\pi\over\beta}~\!n~\!,
  \label{eqn:S1B}
\end{eqnarray}
and, by decomposing into simple fractions,
\begin{eqnarray}
{1\over\beta}\sum_{n\in\mathbb{Z}}{1\over i\omega^F_n-x}~\!{1\over i\omega^F_{n+l}-y}
={1\over\beta}\sum_{n\in\mathbb{Z}}{1\over i\omega^F_n-x}~\!{1\over i\omega^F_n-(y-i\omega_l^B)}
\phantom{aaaaaaa}\nonumber\\
={1\over i\omega^B_l-(y-x)}\left({1\over1+e^{\beta x}}-{1\over1+e^{\beta y}}
\right), \label{eqn:SSF1}\phantom{aa}\\
{1\over\beta}\sum_{n\in\mathbb{Z}}{1\over i\omega^F_n-x}~\!{1\over i\omega^F_{l-n}-y}
=-{1\over\beta}\sum_{n\in\mathbb{Z}}{1\over i\omega^F_n-x}~\!
{1\over i\omega^F_n-(i\omega^B_{l+1}-y)}\phantom{aaaaa}\nonumber\\
={1\over i\omega^B_{l+1}-(y+x)}\left({1\over1+e^{\beta x}}-{1\over1+e^{-\beta y}}
\right). \label{eqn:SSF2}
\end{eqnarray}
Similarly,
\begin{eqnarray}
{1\over\beta}\sum_{l\in\mathbb{Z}}{1\over i\omega^B_l-x}~\!{1\over i\omega^B_l-y}
={1\over x-y}\left({1\over1-e^{\beta x}}-{1\over1-e^{\beta y}}
\right). \label{eqn:SSB1}
\end{eqnarray}
Useful can be also the formula
\begin{eqnarray}
  {1\over x-a_1}~\!{1\over x-a_2}~\!\dots~\!{1\over x-a_n}
  =\sum_{l=1}^n\left(\prod_{k\neq l}^n{1\over a_l-a_k}\right){1\over x-a_l}~\!.
  \label{eqn:SimpleFractions}
\end{eqnarray}

\setcounter{equation}{0}
\renewcommand{\theequation}{B.\arabic{equation}}

\section*{Appendix B}

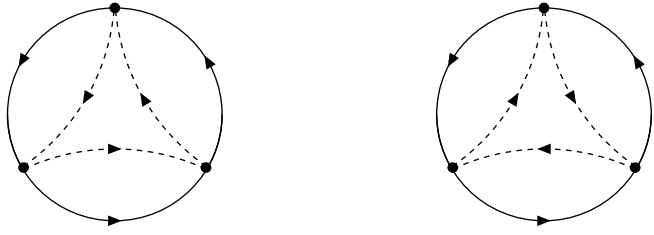
\begin{figure}[]
\begin{center}
\begin{picture}(240,80)(5,0)
  \ArrowArc(40,40)(40,330,90)
  \ArrowArc(40,40)(40,90,210)
  \ArrowArc(40,40)(40,180,360)
  \Vertex(74,20){2}
  \Vertex(6,20){2}
  \Vertex(40,80){2}
  \DashArrowArcn(40,-53)(80,115,65){2}
  \DashArrowArcn(120,86)(80,235,185){2}
  \DashArrowArcn(-40,86)(80,355,305){2}
  \ArrowArc(200,40)(40,330,90)
  \ArrowArc(200,40)(40,90,210)
  \ArrowArc(200,40)(40,180,360)
  \Vertex(234,20){2}
  \Vertex(166,20){2}
  \Vertex(200,80){2}
  \DashArrowArc(200,-53)(80,65,115){2}
  \DashArrowArc(280,86)(80,185,235){2}
  \DashArrowArc(120,86)(80,305,355){2}
\end{picture}
\end{center}
\caption{The particle-particle and the particle-hole diagrams (the
  ``mercedes-like'' diagrams) contributing in the order $C^3_0$ to
  $\Omega^{(3)}$.}
\label{fig:C0cubeMercedes}
\end{figure}

\begin{figure}[]
\begin{center}
\begin{picture}(260,70)(5,0)
  \ArrowLine(20,10)(70,10)
  \ArrowLine(70,10)(45,52)
  \ArrowLine(45,52)(20,10)
  \DashArrowArc(45,60)(8,235,685){2}
  \DashArrowArc(12,7)(8,30,390){2}
  \DashArrowArc(78,7)(8,140,500){2}
  \Vertex(20,10){2}
  \Vertex(70,10){2}
  \Vertex(45,52){2}
  \DashArrowLine(170,10)(220,10){2}
  \DashArrowLine(220,10)(195,52){2}
  \DashArrowLine(195,52)(170,10){2}
  \ArrowArc(195,60)(8,235,685)
  \ArrowArc(162,7)(8,30,390)
  \ArrowArc(228,7)(8,140,500)
  \Vertex(170,10){2}
  \Vertex(220,10){2}
  \Vertex(195,52){2}
\end{picture}
\end{center}
\caption{Two order $C^3_0$, ``mitsubishi-type'' diagrams contributing
  g to $\Omega^{(3)}$.}
\label{fig:C0cube1}
\end{figure}
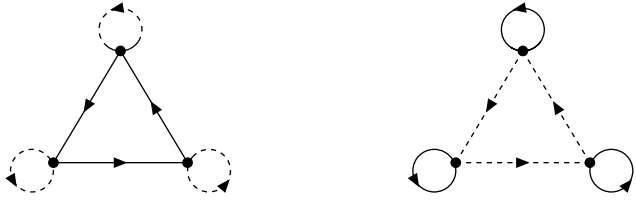

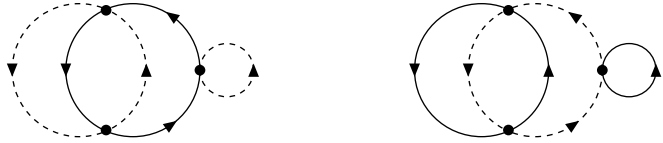
\begin{figure}[]
\begin{center}
\begin{picture}(260,70)(5,0)
\DashArrowArc(30,35)(25,70,290){2}
\DashArrowArc(30,35)(25,290,70){2}
\ArrowArc(50,35)(25,110,250)
\ArrowArc(50,35)(25,250,360)
\ArrowArc(50,35)(25,0,110)
\Vertex(40,12.5){2}
\Vertex(40,57.5){2}
\Vertex(75,35){2}
\DashArrowArc(85,35)(10,180,540){2}
\ArrowArc(180,35)(25,70,290)
\ArrowArc(180,35)(25,290,70)
\DashArrowArc(200,35)(25,110,250){2}
\DashArrowArc(200,35)(25,250,360){2}
\DashArrowArc(200,35)(25,0,110){2}
\Vertex(190,12.5){2}
\Vertex(190,57.5){2}
\Vertex(225,35){2}
\ArrowArc(235,35)(10,180,540)
\end{picture}
\end{center}
\caption{The contributions to $\Omega^{(3)e}$ and $\Omega^{(3)f}$.}
\label{fig:C0cube3}
\end{figure}

Here we demonstrate the cancellation of the additional terms in the
formula
(\ref{eqn:F(3)}) for $F^{(3)}$ against the contributions of diagrams which
vanish at zero temperature. Analogous cancellation of the contribution of
the diagrams shown in Figure \ref{fig:C0squareOther} against the additional
terms in (\ref{eqn:F(2)}) has been checked in the main text.
It will be convenient to introduce first the following notation:
\begin{eqnarray}
  &&n_x\phantom{a}~\!\equiv\int_{\mathbf{k}}\!{1\over1+e^{\beta(\varepsilon_{\mathbf{k}}-x_0)}}~\!,
  \nonumber\\
  &&n_{xx}~\equiv\int_{\mathbf{k}}\!{1\over[1+e^{\beta(\varepsilon_{\mathbf{k}}-x_0)}]^2}~\!,
  \nonumber\\
  &&n_{xxx}\equiv\int_{\mathbf{k}}\!{1\over[1+e^{\beta(\varepsilon_{\mathbf{k}}-x_0)}]^3}~\!,
\nonumber
\end{eqnarray}
etc. From (\ref{eqn:Omega0}) one immediately obtains (all functions are
evaluated at $x_0$ and $y_0$)
\begin{eqnarray}
  &&\Omega^{(0)}_x=-V~\!n_x~\!,\nonumber\\
  &&\Omega^{(0)}_{xx}=-V\beta\left(n_x-n_{xx}\right),\label{eqn:Omega0Derivs}\\
  &&\Omega^{(0)}_{xxx}=-V\beta^2\left(n_x-3n_{xx}+2n_{xxx}\right).\nonumber
\end{eqnarray}
Analogously can be written the derivatives of $\Omega^{(0)}$ with respect to $y$.
The necessary derivatives of $\Omega^{(1)}=C_0V~\!n_x~\!n_y$ take the form
\begin{eqnarray}
  &&\Omega^{(1)}_x=C_0V\beta\left(n_x-n_{xx}\right)n_y~\!,\nonumber\\
  &&\Omega^{(1)}_y=C_0V\beta~\!n_x\left(n_y-n_{yy}\right),\nonumber\\
  &&\Omega^{(1)}_{xx}=C_0V\beta^2\left(n_x-3n_{xx}+2n_{xxx}\right)n_y~\!,
  \label{eqn:Omega1Derivs}\\
  &&\Omega^{(1)}_{yy}=C_0V\beta^2~\!n_x\left(n_y-3n_{yy}+2n_{yyy}\right),\nonumber\\
  &&\Omega^{(1)}_{xy}=C_0V\beta^2\left(n_x-n_{xx}\right)\!\left(n_y-n_{yy}\right).\nonumber
\end{eqnarray}
To $\Omega^{(3)}$, in addition to the two ``mercedes-type'' diagrams
shown in Figure \ref {fig:C0cubeMercedes}
(the contributions $\Omega^{(3)a}$ and $\Omega^{(3)b}$), contribute
also the ``mitsubishi-type'' diagrams of Figure \ref{fig:C0cube1}
(the contributions $\Omega^{(3)c}$ and $\Omega^{(3)d}$),
the two diagrams of Figure \ref{fig:C0cube3} (the contributions
$\Omega^{(3)e}$ and $\Omega^{(3)f}$) and the single ``audi-type''
diagram of Figure \ref{fig:C0cube3} ($\Omega^{(3)g}$).

The computation of $\Omega^{(3)c}$ and $\Omega^{(3)d}$ is straightforward
(it is analogous to that of $\Omega^{(2)b}$ and $\Omega^{(2)c}$) and yields
\begin{eqnarray}
  \Omega^{(3)c}+\Omega^{(3)d}={1\over6}~\!C_0^3V\beta^2\left[
  \left(n_x-3n_{xx}+2n_{xxx}\right)n_y^3+
  n_x^3\left(n_y-3n_{yy}+2n_{yy}\right)\right].\nonumber
\end{eqnarray}
One then readily sees that in (\ref{eqn:F(3)}) this is cancelled by the
last two terms:
\begin{eqnarray}
  \Omega^{(3)a}+\Omega^{(3)b}
  -{\Omega^{(0)}_{xxx}[\Omega^{(1)}_x]^3\over6~\![\Omega^{(0)}_{xx}]^3}
-{\Omega^{(0)}_{yyy}[\Omega^{(1)}_y]^3\over6~\![\Omega^{(0)}_{yy}]^3}=0~\!.\nonumber
\end{eqnarray}

One has now to consider the terms
$-\Omega^{(2)}_x\Omega^{(1)}_x/\Omega^{(0)}_{xx}$ and 
$-\Omega^{(2)}_y\Omega^{(1)}_y/\Omega^{(0)}_{yy}$ in (\ref{eqn:F(3)}).
$\Omega^{(2)}$
is given by three diagrams shown in Figure \ref{fig:ElementaryLoops}
left and \ref{fig:C0squareOther}. It is convenient to write 
the contribution the first one, $\Omega^{(2)a},$ in the form
\begin{eqnarray}
  \Omega^{(2)a}=-{C_0^2V\over2}~\!{1\over\beta}\sum_{l\in\mathbb{Z}}
  \!\int_{\mathbf{q}}\!\int_{\mathbf{k}}\!\int_{\mathbf{p}}
  {1\over\beta}\sum_{n\in\mathbb{Z}}
  {1\over i\omega^F_n-(\varepsilon_{\mathbf{k}}-x)}~\!
  {1\over i\omega^F_{n+l}-(\varepsilon_{\mathbf{k}+\mathbf{q}}-x)}
  \phantom{a}\nonumber\\
  \times{1\over\beta}\sum_{m\in\mathbb{Z}}
  {1\over i\omega^F_m-(\varepsilon_{\mathbf{p}}-y)}
  {1\over i\omega^F_{m-l}-(\varepsilon_{\mathbf{p}-\mathbf{q}}-y)}~\!.
  \label{eqn:C0squareAlternative}
\end{eqnarray}
Differentiating it with respect to $x$ one obtains the expression
which is a sum of the two terms
\begin{eqnarray}
  \Omega^{(2)a}_x={C_0^2V\over2}~\!{1\over\beta}\sum_{l\in\mathbb{Z}}
  \!\int_{\mathbf{q}}\!\int_{\mathbf{k}}\!\int_{\mathbf{p}}
{1\over\beta}\sum_{m\in\mathbb{Z}}
  {1\over i\omega^F_m-(\varepsilon_{\mathbf{p}}-y)}
  {1\over i\omega^F_{m-l}-(\varepsilon_{\mathbf{p}-\mathbf{q}}-y)}
  \phantom{aaaaa}\nonumber\\
  \times{1\over\beta}\sum_{n\in\mathbb{Z}}\left\{
  {1\over[i\omega^F_n-(\varepsilon_{\mathbf{k}}-x)]^2}~\!
  {1\over i\omega^F_{n+l}-(\varepsilon_{\mathbf{k}+\mathbf{q}}-x)}\right.
  \phantom{a}~\nonumber\\
  +\left.{1\over i\omega^F_n-(\varepsilon_{\mathbf{k}}-x)}~\!
  {1\over[i\omega^F_{n+l}-(\varepsilon_{\mathbf{k}+\mathbf{q}}-x)]^2}
  \right\}. \nonumber
\end{eqnarray}
These two terms are equal - to see this, it suffices to set in the second
term $\mathbf{q}=-\mathbf{q}^\prime$, $\mathbf{k}=\mathbf{k}^\prime+\mathbf{q}^\prime$,
$\mathbf{p}=\mathbf{p}^\prime-\mathbf{q}^\prime$ and $l=-l^\prime$, $n=n^\prime+l^\prime$
$m=m^\prime-l^\prime$. Thus, multiplying this by $-\Omega^{(1)}_x/\Omega^{(0)}_{xx}$ which
simply equals $C_0~\!n_y$ one readily sees (taking into account that
${\cal G}_{--}^{(0)}(0,\mathbf{0})=n_y$) that this precisely cancels
$\Omega^{(3)e}$. Thus, in $F^{(3)}$
\begin{eqnarray}
  \Omega^{(3)e}+\Omega^{(3)f}-{\Omega^{(1)}_x\over\Omega^{(0)}_{xx}}~\!\Omega^{(2)a}_x
  -{\Omega^{(1)}_y\over\Omega^{(0)}_{yy}}~\!\Omega^{(2)a}_y=0~\!,\nonumber
\end{eqnarray}

After these cancellations one is left with
\begin{eqnarray}
F^{(3)}=\Omega^{(3)a}+\Omega^{(3)b}+\Omega^{(3)g}
+C_0~\!n_y\left(\Omega^{(2)b}_x+\Omega^{(2)c}_x\right)
+C_0~\!n_x\left(\Omega^{(2)b}_y+\Omega^{(2)c}_y\right)\nonumber\\
+~\!{\Omega^{(1)}_{xx}[\Omega^{(1)}_x]^2\over2~\![\Omega^{(0)}_{xx}]^2}
+{\Omega^{(1)}_{yy}[\Omega^{(1)}_y]^2\over2~\![\Omega^{(0)}_{yy}]^2}
+{\Omega^{(1)}_{xy}\Omega^{(1)}_x\Omega^{(1)}_y\over\Omega^{(0)}_{xx}\Omega^{(0)}_{yy}}~\!.
\phantom{aaaaaaaaaa}\nonumber
\end{eqnarray}
Explicitly the last line of $F^{(3)}$ reads
\begin{eqnarray}
  {1\over2}~\!C_0^3V\beta^2\left\{\left(n_x-3n_{xx}+2n_{xxx}\right)n_y^3\right.
  \phantom{aaaaaaaaaaaaaaaaaaaaaaaaaaa}~\nonumber\\
  \left. +n_x^3\left(n_y-3n_{yy}+2n_{yyy}\right)
  +2~\!n_x\left(n_x-n_{xx}\right)\!\left(n_y-n_{yy}\right)n_y\right\},\nonumber
\end{eqnarray}
while the contribution $\Omega^{(3)g}$ of the ``audi-type''
diagram of Figure \ref{fig:C0cube1} can be written as
\begin{eqnarray}
  \Omega^{(3)g}=C_0^3V\beta^2~\!n_x\left(n_x-n_{xx}\right)\!\left(n_y-n_{yy}\right)n_y~\!.
\nonumber
\end{eqnarray}
Finally,
\begin{eqnarray}
  \Omega^{(2)b}_x+\Omega^{(2)c}_x=-{1\over2}~\!C_0^2V\beta^2\left[
    \left(n_x-3n_{xx}+2n_{xxx}\right)n_y^2+2~\!n_x
    \left(n_x-n_{xx}\right)\!\left(n_y-n_{yy}\right)\right],\nonumber
\end{eqnarray}
(the sum $\Omega^{(2)b}_y+\Omega^{(2)c}_y$ is given by an analogous expression)
and after a straightforward algebra all the extra terms cancel out, so that
eventually, $F^{(3)}=\Omega^{(3)a}+\Omega^{(3)b}$ that is, it given solely
by the ``mercedes-like'' diagrams evaluated at the zeroth order chemical
potentials $x_0$, $y_0$ (or $\tilde x_0$, $\tilde y_0$, if there is an
external magnetic field). The diagrams canceled by the extra terms in the
formulae (\ref{eqn:F(2)}) and  (\ref{eqn:F(3)}) are precisely those (see
e.g. \cite{HamFur00}) which vanish at zero temperature, that is do not
contribute to the expansion of the formula (\ref{eqn:EOmegaFormula}) for the
ground state energy.
\vskip0.2cm

One can also simplify the formula (\ref{eqn:x2Determined}) for the second
order correction $x_2$ to the $\mu_+$ chemical potential (and the analogous
formula for $y_2$). After a straightforward algebra one obtains
\begin{eqnarray}
  x_2=-{\Omega^{(2)a}_x\over\Omega^{(0)}_{xx}}~\!,\nonumber
\end{eqnarray}
all other terms neatly canceling. Of course, $x_1=C_0~\!n_-$,
$y_1=C_0~\!n_+$ but it is perhaps more instructive to write\footnote{This
  form clearly shows, since the cancellation of the divergences in the sum
  $\Omega^{(1)}+\Omega^{(2)a}$ has already been demonstrated, that the
  computed perturbatively chemical potentials $x$ and $y$ are to this
  order finite, after the cutoff dependence of the coupling $C_0$ is
  taken into account. The argument obviously generalizes to all orders:
  if the free energy $F$ is made finite by the renormalization of the
couplings, so must be the chemical potentials.}
\begin{eqnarray}
  x=x_0-{1\over\Omega^{(0)}_{xx}}\left(\Omega^{(1)}_x+\Omega^{(2)a}_x+\dots\right).
  \nonumber
\end{eqnarray}
The cancellation found here is obviously necessary for the
equivalence of two ways of determining the system's polarization.
Indeed, only if this cancellation holds is the minimization of
the free energy written in the form
\begin{eqnarray}
  F=\Omega^{(0)}(x_0,y_0)+(x_0~\!n_x+y_0~\!n_y)~\!V
  +\Omega^{(1)}(x_0,y_0)+\Omega^{(2)a}(x_0,y_0)+\dots, \nonumber
\end{eqnarray}
with respect to $n_x$ (keeping $n_y=n-n_x$), which
(taking into account that $\Omega^{(0)}_x+n_xV=0$, $\Omega^{(0)}_y+n_yV=0$) amounts to
\begin{eqnarray}
(x_0-y_0)~\!V
  =-\left[\Omega^{(1)}_x+\Omega^{(2)a}_x+\dots\right]{\partial x_0\over\partial n_x}
  +\left[\Omega^{(1)}_y+\Omega^{(2)a}_y+\dots\right]{\partial y_0\over\partial n_y}
  ~\!.\nonumber
\end{eqnarray}
equivalent to the condition $\mu_+=\mu_-$ written in the form
$x_0+x_1+x_2+\dots=y_0+y_1+y_2+\dots$, that is
\begin{eqnarray}
  x_0-y_0={\Omega^{(1)}_x+\Omega^{(2)a}_x+\dots\over\Omega^{(0)}_{xx}}
  -{\Omega^{(1)}_y+\Omega^{(2)a}_y+\dots\over\Omega^{(0)}_{yy}}~\!.\nonumber
\end{eqnarray}
The equivalence follows from noticing that since $n_x=-\Omega^{(0)}_x/V$,
$n_y=-\Omega^{(0)}_y/V$, the derivatives of $x_0$ and $y_0$ are precisely equal to
\begin{eqnarray}
  {\partial x_0\over\partial n_x}=-{V\over\Omega^{(0)}_{xx}}~\!,\phantom{aaa}
  {\partial y_0\over\partial n_y}=-{V\over\Omega^{(0)}_{yy}}~\!.\nonumber
\end{eqnarray}
From this argument it immediately follows that
$x_3=-(\Omega^{(3)a}_x+\Omega^{(3)b}_x)/\Omega^{(0)}_{xx}$ and
$y_3=-(\Omega^{(3)a}_y+\Omega^{(3)b}_y)/\Omega^{(0)}_{yy}$.
\vskip0.2cm

\begin{figure}[]
\begin{center}
\begin{picture}(160,40)(5,0)
\ArrowArc(20,35)(20,195,345)
\ArrowArc(20,25)(20,15,165)
\Vertex(40,30){2}
\DashArrowArcn(60,35)(20,345,195){2}
\DashArrowArcn(60,25)(20,165,15){2}
\Vertex(80,30){2}
\ArrowArc(100,35)(20,195,345)
\ArrowArc(100,25)(20,15,165)
\Vertex(120,30){2}
\DashArrowArc(140,35)(20,195,345){2}
\DashArrowArc(140,25)(20,15,165){2}
%
\end{picture}
\end{center}
\caption{The ``audi''-type, order $C^3_0$ contribution to
  $\Omega^{(3)}$.}
\label{fig:AudiType}
\end{figure}
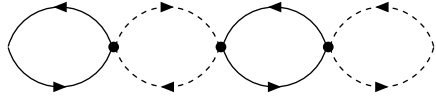

Restricted to the first order the left hand side
of the equality $x_0+x_1=y_0+y_1$ reads
\begin{eqnarray}
  x_0+x_1+\dots=k_{\rm B}T~\!f^{-1}\!\left(\left(
  {\varepsilon^{(0)}_{\rm F}(n_+)\over k_{\rm B}T}\right)^{3/2}\right)
  +C_0~\!n_-+\dots\nonumber
\end{eqnarray}
The right hand side is given by the analogous formula. If one sets
here $n_\pm=(N/2V)(1\pm P)$, this reproduces
the condition (\ref{eqn:KesioCond}).

\end{document}